\documentclass[aps,reprint,groupedaddress]{revtex4-2}

\usepackage{amsmath}
\usepackage{amssymb}
\usepackage{graphicx}
 \usepackage{epstopdf}
\usepackage{float}

\usepackage{placeins}
\usepackage{mathtools}
\usepackage{bm}
\usepackage{array,multirow}

\usepackage{xcolor}
\usepackage{booktabs}

\usepackage{chngcntr}

\usepackage[normalem]{ulem}

\begin{document}

\title{Big jump principle for heavy-tailed random walks with correlated increments}

\author{Marc H\"oll, Eli Barkai}

\affiliation{Department of Physics, Institute of Nanotechnology and Advanced Materials, Bar-Ilan University, Ramat-Gan 52900, Israel}

\begin{abstract}
The big jump principle explains the emergence of extreme events for physical quantities modelled by a sum of independent and identically distributed random variables which are heavy-tailed. Extreme events are large values of the sum and they are solely dominated by the largest summand called the big jump. Recently, the principle was introduced into physical sciences where systems usually exhibit correlations. Here, we study the principle for a random walk with correlated increments. Examples are the autoregressive model of first order and the discretized Ornstein-Uhlenbeck process both with heavy-tailed noise. The correlation leads to the dependence of large values of the sum not only on the big jump but also on the following increments. We describe this behaviour by two big jump principles, namely unconditioned and conditioned on the step number when the big jump occurs. The unconditional big jump principle is described by a correlation dependent shift between the sum and maximum distribution tails. For the conditional big jump principle, the shift depends also on the step number of the big jump. 
\end{abstract}

\maketitle

\section{Introduction}

The big jump principle (BJP) is a concept of extreme value statistics. Since its first formulation in 1964 \cite{chistyakov1964theorem}, studies are ongoing in fields like large deviation theory \cite{mikosch2016large,
konstantinides2005large,
mikosch2000supremum} and financial mathematics \cite{embrechts1982estimates,
rolski2009stochastic,
kyprianou2006introductory}. The original principle \cite{chistyakov1964theorem} relates the sum $x_N=\sum_{i=1}^N \delta_i$ and the maximum $\delta_\text{max}=\text{max}(\delta_1,\ldots,\delta_N)$ of $N$ independent and identically distributed (IID) random variables $(\delta_1,\delta_2,\ldots,\delta_N)$ following a subexponential (e.g. heavy-tail) distribution. The principle states that the tail distributions are asymptotically equal
\begin{equation}\label{bjpintro}
\text{Prob}(x_N>z)\sim\text{Prob}(\delta_\text{max}>z)
\end{equation}
for large $z$ and any finite $N$. This formula means that the ratio of both sides converges $\text{Prob}(x_N>z) / \text{Prob}(\delta_\text{max}>z) \to 1$ for $z\to\infty$, i.e. the right tails of the distributions match. Large values of the sum $x_N$ are solely described by large values of maximum $\delta_\text{max}$. The latter is called the big jump. All remaining summands are negligible because the big jump dominates the statistics of the sum. \\ \\

In the language of physics, the BJP connects large excursion of a random walk model $x_N$ with the maximum increment $\delta_\text{max}$. The principle explains the emergence of macroscopical extreme events by the emergence of a single microscopical extreme event. Roughly speaking, this scenario is very different from what we find for  processes modelled with narrow distributions, where the extreme of the sum is controlled by the accumulation of many small jumps all adding up to the extreme event (a case which is typically exponentially unlikely). The main challenge is the corporation of correlations into the theory. Recently, the BJP draw attention in statistical physics \cite{vezzani2019single,
wang2019transport,
burioni2020rare,
vezzani2020rare,
wang2020large}. These works studied, 
amongst other things, L\'evy walks, renewal processes, continuous time random walks and the L\'evy Lorentz gas. Other than the BJP, the statistics of extreme events are studied for many different correlated physical systems \cite{MAJUMDAR20201,
godreche2017longestaa,
godreche2021condensation,
holl2020extreme,
bar2016exact,
mori2021detecting,
grebenkov2021exact}. \\ \\


This article is a new addition for the investigation of the BJP in physical systems where correlations are present. We demonstrate an exactly solvable model showing the detailed influence of the correlations on the big jump principle. We study a random walk model $\tilde{x}_N = \sum_{i=1}^N \tilde{\delta}_i$ with correlated increments $\tilde{\delta}_i$. The correlated process is constructed via increments given by the weighted sum $\tilde{\delta}_i=\sum_{j=1}^i M_{i-j} \delta_j$ with some positive memory kernel $M_{i-j}$. Here as before the $\delta_j$s are IID subexponential random variables following the BJP Eq.~(\ref{bjpintro}). This model appears in many different contexts. An import example are waiting times (or steps) appearing in the correlated continuous time random walk \cite{meerschaert2009correlated,
chechkin2009continuous,
schulz2013correlated,
tejedor2010anomalous,
montero2007nonindependent,
comolli2018impact} with application in animal telemetry data \cite{johnson2008continuous}, biological movements \cite{maye2007order}, external force fields \cite{magdziarz2012correlated} and porous media diffusion \cite{de2013flow}. This problem can also be seen as the discrete Ornstein-Uhlenbeck process with heavy-tailed noise \cite{janczura2011subordinated, fink2011fractional} and the autoregressive model of first order with the corresponding heavy-tailed noise \cite{mikosch1995parameter,
liu2019parameter,
samoradnitsky2017stable,
embrechts2013modelling} which is a specific example of fractional L{\'e}vy stable motion \cite{burnecki2010fractional,
burnecki2017identification,
dybiec2006levy,
dybiec2007escape}. \\ \\

We will find two ways to extend the BJP, i.e. we give the relationship between statistics of large values of the sum $\tilde{x}_N$ and of large values of the maximum increment $\tilde{\delta}_\text{max}=\text{max}(\tilde{\delta}_1,\ldots,\tilde{\delta}_N)$.  Let $b$ be the step number when we locate the maximum of $\tilde{\delta}_\text{max}=\tilde{\delta}_b$. In the first method, we condition on $b$. The second method is unconditioned on the step  number of the maximum occurrence. In contrast to the IID case, the step number $b$ is relevant. Large values of the sum $\tilde{x}_N$ are described not only by the big jump but also by the subsequent increments following the big jump (see conditions on the memory kernel below). Of course, this is due to the effects of correlations with positive $M_{i-j}$. For both cases, we find a renormalization of the big jump, taking the sequence of all subsequent increments after the big jump into account.\\ \\

The article is constructed as follows. Sec.~\ref{sec:random} introduces the model. Sec.~\ref{sec:main} presents the main results. Sec.~\ref{sec:example} discusses an example with $N=2$ increments. Sec.~\ref{sec:uncond} derives the unconditional BJP. Sec.~\ref{sec:cond} derives the conditional BJP. Sec. \ref{sec:outlook} give a summary.

\section{Random Walk with Correlated Increments}\label{sec:random}

\subsection{Basics}

We consider the IID random variables $(\delta_1,\ldots,\delta_N)$ with a probability density function (PDF) following a power law
\begin{equation}\label{pdfpowerlaw}
f_{\delta_i}(z) \sim A z^{-1-\alpha}
\end{equation}
for large $z$. The notation $f_{\delta_i}(z)$ indicates that the random variable is written as the subscript and $z$ is its value. The exponent is $0<\alpha<2$ with $\alpha \neq 1$ and $A$ is some prefactor. Then the IID BJP reads $f_{x_N}(z) \sim f_{\delta_\text{max}}(z) \sim NA z^{-1-\alpha}$. This relation is valid for any $N$. A similar notation is $x_N \stackrel{d}{\sim} \delta_\text{max}$ which means that both random variables are similarly distributed for large values. In addition, in the large $N$ limit the IID BJP connects  the L\'evy  generalized central limit theorem (dealing with the sum of random variables) with the tail of the Fr\'echet distribution (dealing with the maximum).\\ \\

In this article, we consider as examples one-sided and two-sided PDFs. As an example for a one-sided PDF, we investigate the Pareto PDF $f_{\delta_i}(z)=\alpha z^{-1-\alpha}$ with $z \ge 1$ and otherwise this function is equal zero. The mean is $\langle \delta_i \rangle = \int_1^\infty z f_{\delta_i}(z) \mathrm{d}z=\alpha/(\alpha-1)$ for $\alpha\in(1,2)$ while no mean exists for $\alpha\in(0,1)$. As a two-sided PDF, we consider the symmetric $\alpha$-stable PDF $f_{\delta_i}(z)=L_{\alpha,\kappa=0,c=1,\mu=0}(z)$ with the sknewness parameter $\kappa=0$, the scale parameter $c=1$ and the location parameter $\mu=0$. In general, the L\'evy stable PDF $L_{\alpha,\kappa,c,\mu}(z)$ is defined via the characteristic function $\varphi(k)=\langle\text{exp}(ikz)\rangle=\text{exp}\left[ik\mu - c |k|^\alpha \left(1-i\kappa\text{sign}(k)\text{tan}\left(\pi\alpha/2\right)\right)\right]$ which is for the symmetric PDF given by $\varphi_{\delta_i}(k) = \text{exp}(-|k|^\alpha)$. The large $z$ behaviour is $f_{\delta_i}(z) \sim A |z|^{-1-\alpha}$ with $A=\text{sin}(\pi\alpha/2) \Gamma(\alpha+1)/\pi$. \\ \\

In App.~\ref{sec:correction}, we present a way to find corrections of the IID BJP for the Pareto random variables with $\alpha>1$, i.e. the mean $\langle \delta_i \rangle$ exists and is positive. The basic idea is to replace the remaining variables (not being the maximum) by their mean $\langle \delta_i \rangle$, i.e. we neglect the fluctuations of the remaining random variables. Then we get the correction to the IID BJP as $x_N \stackrel{d}{\sim} \delta_\text{max}  + (N-1) \langle \delta_i \rangle$. In Fig.~\ref{fig:plot_app_01}, we see that this correction describes the sum distribution pretty well also left from the right tail. We present in App.~\ref{sec:correction} the derivation

\subsection{Model}

The crucial assumption of the BJP in Eq.~(\ref{bjpintro}) is the IID behaviour of the increments. We will drop this assumption now. We consider the random walk model
\begin{equation}
\tilde{x}_N = \sum_{i=1}^N \tilde{\delta}_i
\end{equation}
with correlated increments. We construct the increments by a sum of IID heavy-tailed random variables weighted with a memory kernel
\begin{equation}\label{inccorr}
\tilde{\delta}_i=\sum_{j=1}^i M_{i-j} \delta_j
\end{equation}
where the kernel is positive $M_{i-j}>0$. The kernel is decreasing $M_{i-j_2}<M_{i-j_1}$ for increasing steps $j_2>j_1$ and we define $M_0=1$. Here as before, the IID random variables $\delta_i$ are given by L\'evy or Pareto PDFs though the theory is actually more general. \\ \\

This type of a correlated process was studied,  inter alia, in \cite{chechkin2009continuous} for the continuous time random walk model with correlated waiting times. An illustrative and important example is given by the memory kernel
\begin{equation}\label{kernelar}
M_{i-j}=m^{i-j}
\end{equation}
where the parameter $0<m<1$ controls the correlations. We call this the exponential kernel because $m^{i-j}=\text{exp}[(i-j)\text{log}(m)]$ with the correlation length $-1/\text{log}(m)$. The increments follow the recursive equation
\begin{equation}
\tilde{\delta}_i = m \tilde{\delta}_{i-1} + \delta_i.
\end{equation}
This formula can be found in different fields. For example, it is the discretized version of the heavy-tailed Ornstein-Uhlenbeck process (when $\delta_i$ follows the symmetric $\alpha$-stable PDF). In the field of time series analysis, this formula is also known as autoregressive model of first order with heavy-tailed noise \cite{liu2019parameter,
samoradnitsky2017stable,
embrechts2013modelling}. Another example for the memory kernel in Eq.~(\ref{inccorr}) is the power law kernel
\begin{equation}\label{kernellrc}
M_{i-j}=(i-j+1)^{-\beta}
\end{equation}
with $0<\beta<1$. We call this the algebraical kernel. This kernel is non-integrable in contrast to Eq.~(\ref{kernelar}).\\ \\

We rearrange the summands of the random walker $\tilde{x}_N=\sum_{i=1}^N\left(\sum_{j=1}^i M_{i-j}\delta_j\right)$ with respect to the IID random variables $\delta_i$ and obtain
\begin{equation}\label{procorr}
\tilde{x}_N = \sum_{k=1}^N W_{N-k}\delta_k
\end{equation}
with the weight
\begin{equation}\label{weights}
W_{N-k}=\sum_{l=0}^{N-k}M_l.
\end{equation}
We changed the indices to indicate this rearrangement. For the exponential memory kernel $M_l=m^l$ the weight is
\begin{equation}\label{anotherlarge}
W_{N-k} = \frac{1-m^{N-k+1}}{1-m} \sim \frac{1}{1-m}
\end{equation}
where we used $0<m<1$ in the large $N$ limit with fixed $k$. On the other hand, the large $N$ limit with fixed $k$ of the weight for the algebraical kernel $M_l=(l+1)^{-\beta}$ is
\begin{equation}\label{weightalg}
W_{N-k}\sim\frac{N^{1-\beta}}{1-\beta}.
\end{equation}
Since $0<\beta<1$, the weight $W_{N-k}$ is increasing with $N$ indicating a stronger correlation compared to the exponential memory case.

\subsection{Limiting PDF of $\bm{\tilde{x}_N}$ for large $\bm{N}$}

Before we continue with the derivation of the BJP, we present the limiting distribution of $\tilde{x}_N$ for large $N$ in order to get a first feeling for the basic properties of the studied process. For the sum of uncorrelated increments, the limiting distribution is stable as described by L\'evy's generalised central limit theorem. The question we address is how this changes in the presence of correlations. For study of correlated sums, we also refer to \cite{hilhorst2009central,
bertin2006generalized,
baldovin2007central}. \\ \\ 

Let's begin with the two-sided symmetric PDF $f_{\delta_i}(z)=L_{\alpha,0,1,0}(z)$. Since $\tilde{x}_N$ is a linear combination of L\'evy distributed random variables, the sum is also described by this law with a new scale parameter \begin{equation}\label{largenlimit1}
f_{\tilde{x}_N}(z) = L_{\alpha,0,c,0}(z)
\end{equation}
which is valid for any $N$. See the definition of $L_{\alpha,\kappa,c,\mu}(z)$ above. Here, $c=\tilde{\gamma}_N$ is given by the scale factor
\begin{equation}\label{scalingfa}
\tilde{\gamma}_N = \sum_{k=1}^N (W_{N-k})^\alpha.
\end{equation}
In the case of no correlation, i.e. $M_0=1$ and $M_l=0$ for $l\ge 1$, this factor is $\tilde{\gamma}_N=N$. An important observation is the large $N$ behaviour of $\tilde{\gamma}_N$ for the two different memory kernels which we derive in App.~\ref{sec:scaling}. For the exponential memory kernel it scales linearly
\begin{equation}
\tilde{\gamma}_N \sim \frac{1}{(1-m)^\alpha} N.
\end{equation}
We see that this is the scaling of the IID case, besides a renormalized prefactor. For the algebraical memory kernel the scaling factor behaves as
\begin{equation}
\tilde{\gamma}_N \sim \frac{1}{(1-\beta)^\alpha} \frac{N^{1+\alpha(1-\beta)}}{1+\alpha(1-\beta)}
\end{equation}
with the exponent $1<1+\alpha(1-\beta)<3$. \\ \\

On the other hand, the sum $\tilde{x}_N$ for the Pareto PDF $f_{\delta_i}(z)=\alpha z^{-1-\alpha}$, $z \ge 1$, is described by a L\'evy stable law only in the large $N$ limit, similar to the IID case. We show in App.~\ref{sec:applargenpdf} that the PDF behaves as
\begin{equation}\label{largenlimit2}
f_{\tilde{x}_N}(z) \sim L_{\alpha,\kappa,c,\mu}(z)
\end{equation}
for large $N$. The skewness parameter is $\kappa=1$. The scale parameter is $c=-\alpha\Gamma(-\alpha) \text{cos}\left(\pi\alpha/2\right) \tilde{\gamma}_N$ where the scaling factor $\tilde{\gamma}_N$ is defined as in Eq.~(\ref{scalingfa}). The location parameter is $\mu=0$ for $\alpha\in(0,1)$ and $\mu=\langle \tilde{x}_N\rangle=\langle \delta_i \rangle \tilde{\lambda}_N$ for $\alpha\in(1,2)$ with the location factor
\begin{equation}
\tilde{\lambda}_N=\sum_{k=1}^N W_{N-k}.
\end{equation}
Note that $\tilde{\lambda}_N=\tilde{\gamma}_N$ for $\alpha=1$. Analogue to the procedure in App.~\ref{sec:applargenpdf} we can derive the large $N$ behaviour of the location factor. For the exponential memory kernel, it behaves as
\begin{equation}\label{seq01}
\tilde{\lambda}_N \sim \frac{1}{1-m} N .
\end{equation}
For the algebraical memory kernel, the location parameter behaves as
\begin{equation}\label{seq02}
\tilde{\lambda}_N \sim \frac{1}{(1-\beta)(2-\beta)} N^{2-\beta} 
\end{equation}
with the exponent $1<2-\beta<2$. In case of no correlation, the location factor is for any $N$ given by $\tilde{\lambda}_N=N$.\\ \\

Let us conclude this subsection, with two observations. First, the large $z$ behaviour of right hand sides of Eq.~(\ref{largenlimit1}) and (\ref{largenlimit2}) is $f_{\tilde{x}_N}(z) \sim \tilde{\gamma}_N Az^{-1-\alpha}$. Note that here the symbol $\sim$ means also large $N$ for the Pareto case, see Eq.~(\ref{largenlimit2}) (we show later that this large $z$ behaviour is also valid for finite $N$). Secondly, for large $N$, the PDF of the rescaled correlated sum $(\tilde{\gamma}_N)^{-1/\alpha}(\tilde{x}_N-\mu)$ converges to the $N$-independent L\'evy PDF $L_{\alpha,\kappa,c/\tilde{\gamma}_N,0}(z)$ where $\kappa$ and $c$ are stated as above and the large $z$ behaviour is $A z^{-1-\alpha}$. In Fig.~\ref{fig:plot_17} in App.~\ref{sec:applargenpdf} we show this limiting behaviour. This result can be seen as a modification of the generalised central limit with a renormalization of the scaling sequences Eq.~(\ref{seq01}) and (\ref{seq02}). In the following, we leave the $N\to\infty$ case and derive the BJPs for any finite $N$ where the only effect is renormalisation of the parameters. Importantly, in this section, we learned how the memory kernel leads to a renormalisation effect described by $\tilde{\gamma}_N$ and $\tilde{\lambda}_N$. We will find a similar behaviour again when treating the BJPs.

\section{Main Results}\label{sec:main}

\begin{figure}\begin{center}
\includegraphics[width=0.23\textwidth]{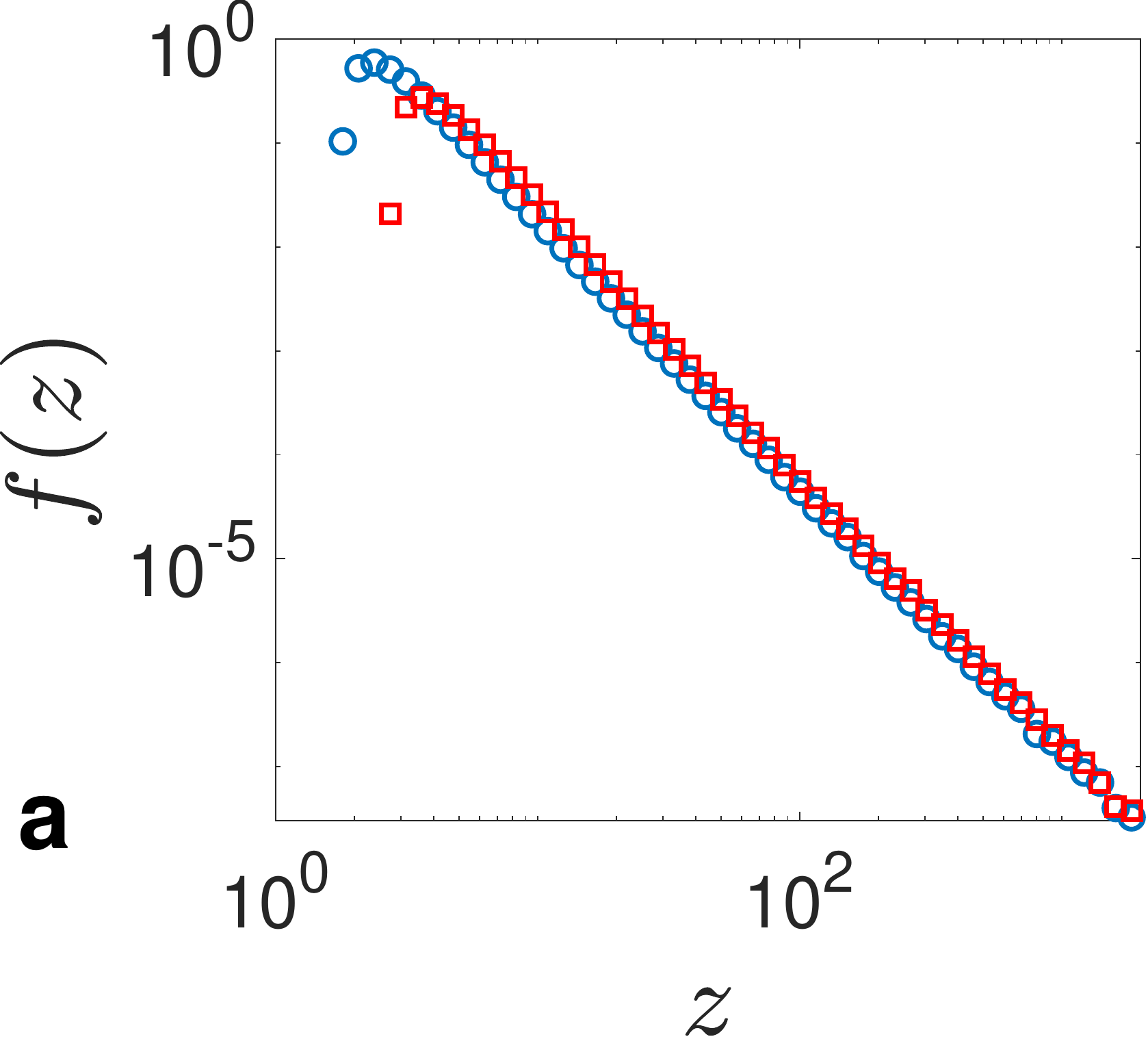}
\includegraphics[width=0.23\textwidth]{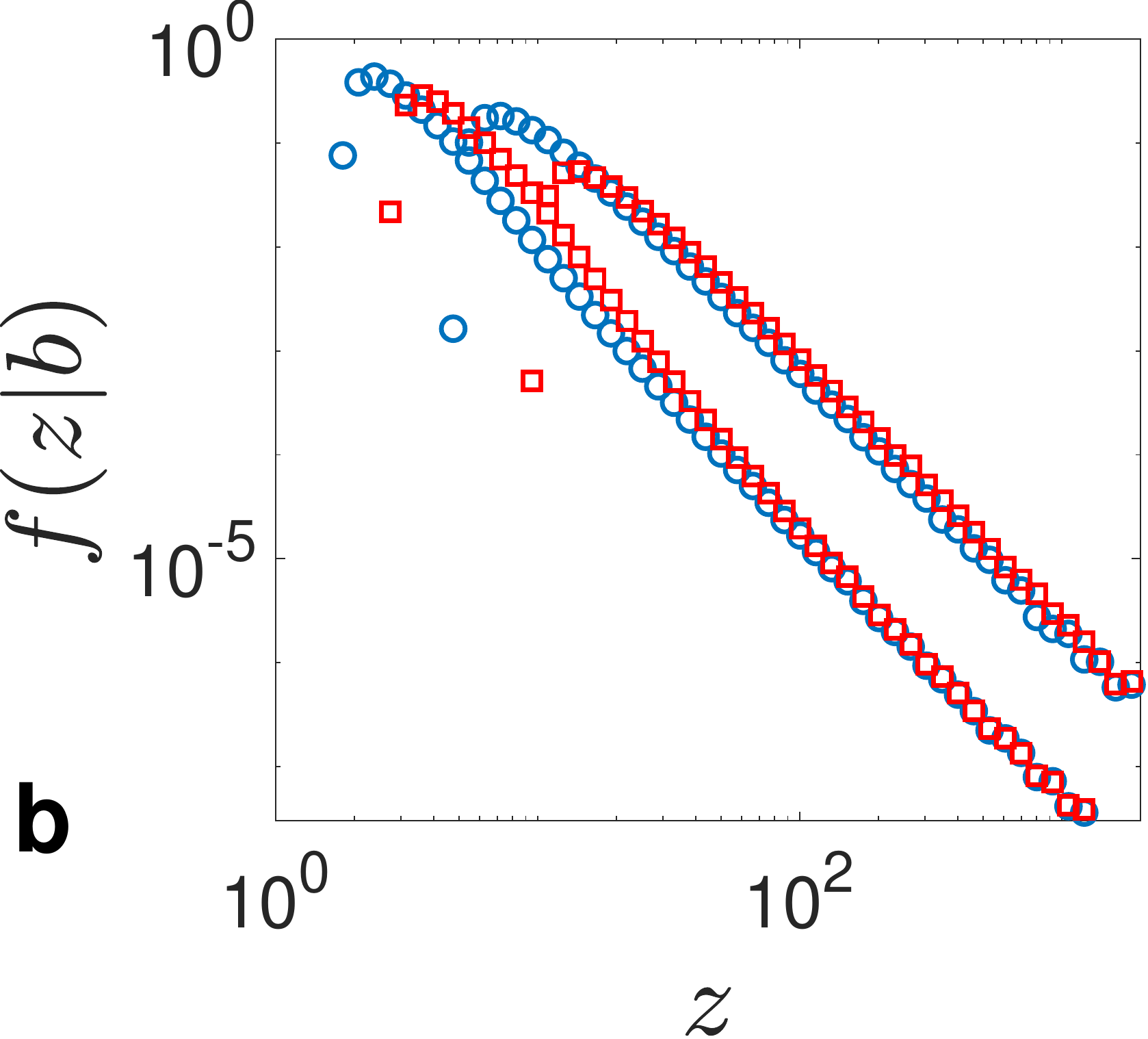}
\caption{\textbf{a} The unconditional BJP for $N=2$ with the memory kernel $M_1=0.8$. The figure shows the histograms of the rescaled maximum $(\tilde{\gamma}_N/N)^{1/\alpha}\tilde{\delta}_\text{max}$ (blue circles) and the sum $\tilde{x}_N$ (red squares). These unconditional distributions match as predicted in Eq.~(\ref{unbjp1}). We used $\delta_i$ following the Pareto PDF with $\alpha=1.5$ and $10^7$ realizations. \textbf{b} The conditional BJP with the same parameters as in figure a. The figure shows the rescaled conditional maximum $W_{N-b}\tilde{\delta}_\text{max}$ (blue circles) and the conditional sum $\tilde{x}_N$ (red squares) both conditioned on the maximum step number $b=1$ (the two upper curves) and $b=2$ (the two lower curves). The conditional distributions match as predicted in Eq.~(\ref{cobjp2}). \label{fig:fig_01-02_really_new}}
\end{center}\end{figure}

We present the two main results of this article. As mentioned, we want to study the relationship between the distribution of the sum $\tilde{x}_N$ and of the maximum correlated increment
\begin{equation}\label{naivemax}
\tilde{\delta}_\text{max}=\text{max}(\tilde{\delta}_1,\tilde{\delta}_2,\ldots,\tilde{\delta}_N),
\end{equation}
i.e. what is the BJP for the correlated random walk $\tilde{x}_N$? Instead of a unique BJP as in the IID case Eq.~(\ref{bjpintro}), the random walk model with correlated increments exhibits two BJPs. We call them the unconditional and the conditional BJP. Let the maximum $\tilde{\delta}_\text{max}$ happen at the step number $b=1,\ldots,N$, i.e. $\tilde{\delta}_\text{max}=\tilde{\delta}_b$. The unconditional BJP is
\begin{equation}\label{unbjp1}
\text{Prob}(\tilde{x}_N>z) \sim \frac{\tilde{\gamma}_N}{N} \text{Prob}\left(\tilde{\delta}_\text{max}>z\right)
\end{equation}
for large $z$ and any $N$. This relationship is independent of $b$. The conditional BJP is
\begin{equation}\label{cobjp2}
\text{Prob}(\tilde{x}_N>z|b) \sim  (W_{N-b})^{\alpha} \text{Prob}\left(\tilde{\delta}_\text{max}>z|b\right)
\end{equation}
conditioned on the occurrence of the big jump $\tilde{\delta}_\text{max}$ at $b$ for large $z$ and any $N$. We show in Fig.~\ref{fig:fig_01-02_really_new} these two BJPs for the model with $N=2$ and observe perfect matching as predicted by the theory. Before we derive these results in generality (i.e. for general $N$), we discuss this case $N=2$ next to understand the nature of both BJPs.

\section{Example with \bm{$N=2$}}\label{sec:example}

\begin{figure}\begin{center}
\includegraphics[width=0.47\textwidth]{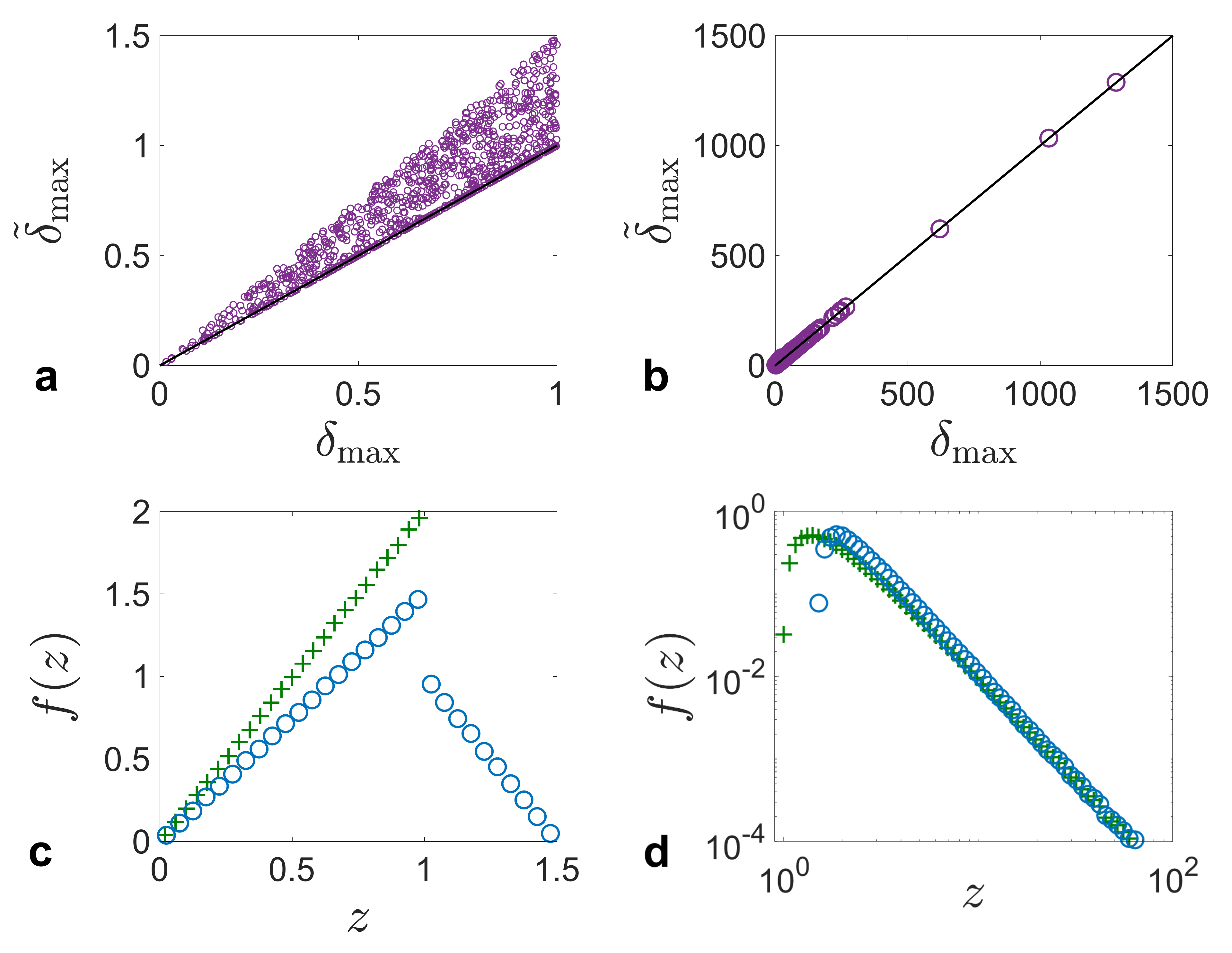}

\caption{\textbf{a} Scatter plot between $\delta_\text{max}$ and $\tilde{\delta}_\text{max}$ (purple circles) where $N=2$. The $\delta_i$ follow the uniform PDF on the interval $[0,1]$. The memory kernel is $M_0=1$ and $M_1=0.5$. The number of realizations is $10^3$. A line with slope one is shown (black line). There is no linear relationship between $\delta_\text{max}$ and $\tilde{\delta}_\text{max}$ when both are large since the parent distribution is narrow. \textbf{b} The same parameters as the left scatter plot but now the IID random variables $\delta_i$ follow the Pareto PDF with $\alpha=3/2$. There is a linear relationship between $\delta_\text{max}$ and $\tilde{\delta}_\text{max}$ when both are large as predicted in Eq.~(\ref{correlatedmaximu}) (black line). While the naked eye sees in this example a linear relationship for large values (on the right) we also obtained the correlation coefficient $r_c$ using the top $1\%$ of the data set (see main text). It is $0.119$ for the sample in the left and $0.999$ for the right scatter plot. \textbf{c} Histograms of $\delta_\text{max}$ (green crosses) and $\tilde{\delta}_\text{max}$ (blue circles) for the left scatter plot with the uniform IID random variables but now with $10^6$ realizations. There is no matching of the distributions. \textbf{d} Histograms of $\delta_\text{max}$ (green crosses) and $\tilde{\delta}_\text{max}$ (blue circles) for the right scatter plot with the Pareto IID random variables but now with $10^6$ realizations. There is a matching of the right tails.\label{fig:plot_01-04}}
\end{center}\end{figure}

\begin{figure}\begin{center}
\includegraphics[width=0.45\textwidth]{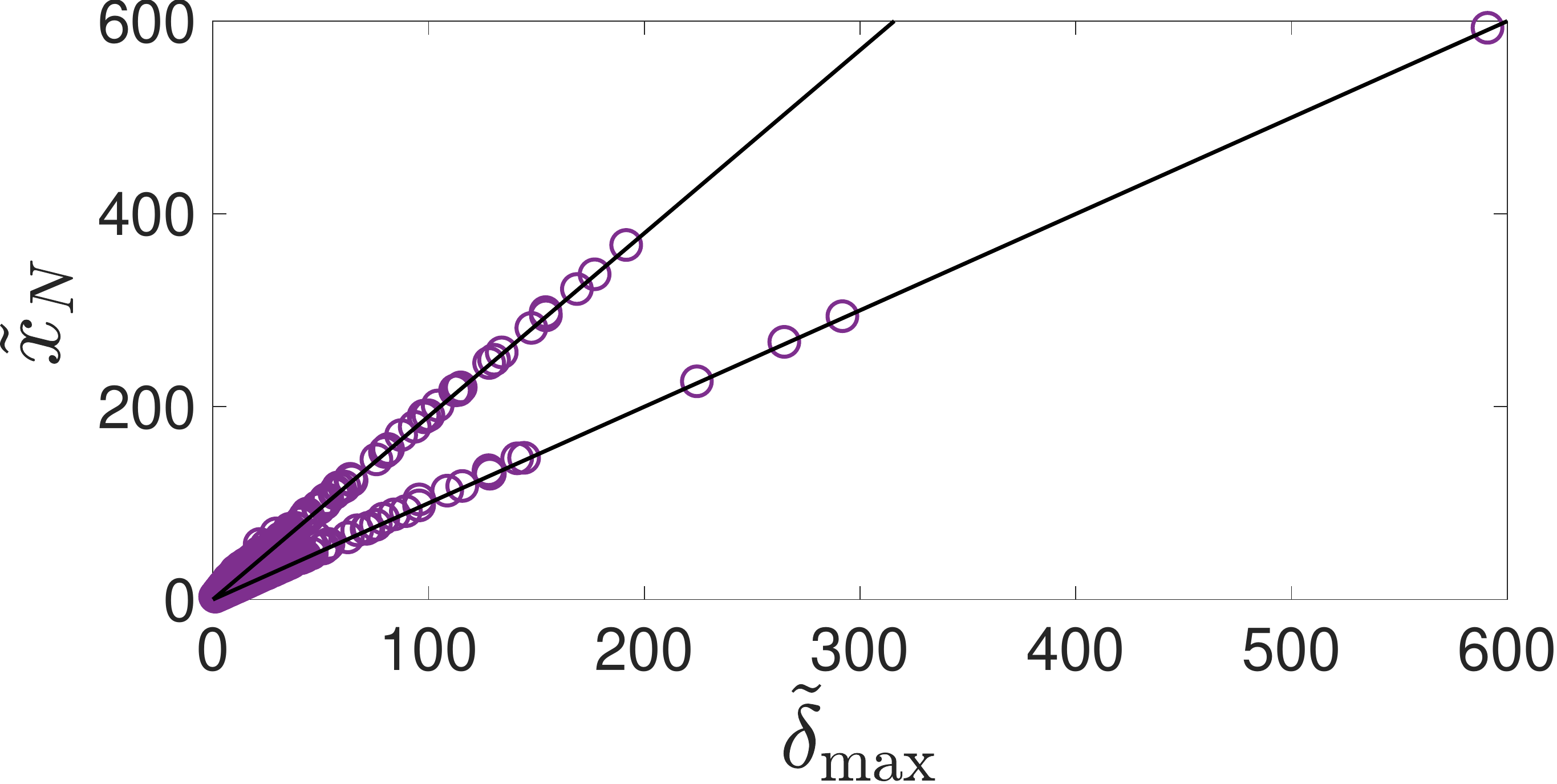}
\caption{Scatter plot between $\tilde{\delta}_\text{max}$ and $\tilde{x}_N$ (purple circles) where $N=2$. The $\delta_i$ follow the Pareto PDF with $\alpha=3/2$. The memory kernel is exponential with $m=0.9$. The number of realizations is $10^4$. The two lines have the slope $W_0=1$ and $W_1=1+m$ (black lines). The two linear relationships between $\tilde{\delta}_\text{max}$ and $\tilde{x}_N$ when both are large are predicted in Eq.~(\ref{correlatedbigju2}) and indicate at which step $\tilde{\delta}_\text{max}$ occurred. The lower line with slope $W_0$ are processes where the big jump happened at $b=2$ while the line with slope $W_1$ means $b=1$.\label{fig:plot_05}}
\end{center}\end{figure}

In this section, we present the key points for $N=2$. All statements in this section will be derived later in generality for any $N$. So, we consider the two increments 
\begin{equation}\begin{split}
\tilde{\delta}_1 &= \delta_1, \\
\tilde{\delta}_2 &= M_1\delta_1 + \delta_2
\end{split}\end{equation}
and the random walk is found on
\begin{equation}
\tilde{x}_2 = \tilde{\delta}_1 + \tilde{\delta}_2 =  \delta_1 + M_1\delta_1 + \delta_2
\end{equation}
after two steps. Recall that $M_1<1$ and this is valid for both the exponential and algebraic memory case. The IID random variables $\delta_1$ and $\delta_2$ are heavy-tail distributed. We have to discuss two questions: What is the influence of the maximum of the IID random variables $\delta_\text{max}=\text{max}(\delta_1,\delta_2)$, first, on the big jump $\tilde{\delta}_\text{max}=\text{max}(\tilde{\delta}_1,\tilde{\delta}_2)$ and, secondly, on the correlated random walker $\tilde{x}_2$? The answers to both questions will enable us to understand the relationship between $\tilde{\delta}_\text{max}$ and $\tilde{x}_2$.\\ \\

The BJP for the IID heavy-tailed random variables $\delta_1$ and $\delta_2$ of Eq.~(\ref{bjpintro}) is $\delta_1+\delta_2 \stackrel{d}{\sim} \delta_\text{max}$. We always compare the right tails of the distributions. This identity holds in the limit when the random variable $\delta_\text{max}$ is very large.  Importantly, it is irrelevant for this IID BJP which of the two random variables $\delta_1$ and $\delta_2$ is the maximum. \\ \\

Assuming the maximum of the IID random variables is the first one $\delta_\text{max}=\delta_1$. The increments are $\tilde{\delta}_1=\delta_\text{max}$ and $\tilde{\delta}_2=M_1\delta_\text{max}+\delta_2$. Further assuming that $\delta_\text{max}$ is very large reduces the second increment to only one term $\tilde{\delta}_2 \approx M_1 \delta_\text{max}$. This term is smaller than $\delta_\text{max}$ since $M_1<1$ so that the big jump is $\tilde{\delta}_\text{max}=\tilde{\delta}_1=\delta_\text{max}$. On the other hand, assuming the maximum of the IID random variables is the second one $\delta_\text{max}=\delta_2$. Then the increments are $\tilde{\delta}_1=\delta_1$ and $\tilde{\delta}_2=M_1\delta_1+\delta_\text{max}$. Here, it is $\tilde{\delta}_2 > \tilde{\delta}_1$ so that $\tilde{\delta}_\text{max}=\tilde{\delta}_2$. Further assuming that $\delta_\text{max}$ is very large allows to neglect the term $M_1\delta_1$ in $\tilde{\delta}_2$ so that $\tilde{\delta}_\text{max} \approx\delta_\text{max}$. Both cases are summarized with the observation that the big jump is approximately the same as the maximum of the IID random variables
\begin{equation}\label{correlatedmaximu}
\tilde{\delta}_\text{max} \simeq \delta_\text{max}
\end{equation}
when both are large enough, see the scatter plot in Fig.~\ref{fig:plot_01-04}. We present two examples, one where Eq.~(\ref{correlatedmaximu}) doesn't hold, that is the uniforum PDF for $\delta_i$, and one example where the equation holds, that is the Pareto PDF for $\delta_i$. In App.~\ref{sec:pareto}, we show Eq.~(\ref{correlatedmaximu}) by calculating the exact expression of the PDF of $\tilde{\delta}_\text{max}$ and then approximate for large values.\\ \\

A relationship as Eq.~(\ref{correlatedmaximu}) yields a straight line in the scatter plot for large values which is a first visible indicator of the existence of such big jump principle relationships. In order to verify this behaviour we suggest using the Pearson correlation coefficient between the samples of $\tilde{\delta}_\text{max}$ and $\delta_\text{max}$, namely $r_c=\langle (\tilde{\delta}_\text{max}-\langle \tilde{\delta}_\text{max}\rangle)(\delta_\text{max}-\langle \delta_\text{max} \rangle)\rangle/[\sqrt{\langle \tilde{\delta}_\text{max}^2\rangle-\langle \tilde{\delta}_\text{max}\rangle^2}\sqrt{\langle \delta_\text{max}^2\rangle-\langle \delta_\text{max}\rangle^2}]$ where the averaging $\langle \ldots \rangle$ is performed over the sample. We suggest the test a particular number of large values, let's say one percent of the largest $\delta_\text{max}$. Then for the uniform $\delta_i$ in Fig.~\ref{fig:plot_01-04}, we find $r_c\approx 0.119$ and for the Pareto $\delta_i$, we find $r_c\approx 0.999$. A value near one $r_c\approx 1$ indicates that there is indeed a strong linear relationship between large values of $\tilde{\delta}_\text{max}$ and $\delta_\text{max}$.\\ \\

 We also show the PDFs of $\delta_\text{max}$ and $\tilde{\delta}_\text{max}$ in Fig.~\ref{fig:plot_01-04}. For the uniform random variables, there is no matching of the PDFs in the rights tails. In particular, we have $f_{\delta_\text{max}}(z)=2z$ for $0<z<1$ and, on the other hand, $f_{\tilde{\delta}_\text{max}}(z)=(2-M_1)z$ for $0<z<1$ and $(1+M_1-z)/M_1$ for $1<z<1+M_1$ which is derived in App.~\ref{sec:uniform}. Both PDFs are clearly different. In case of the Pareto PDF, we observe that the right tails of the maximum distributions match. Below, we will focus on the matching of the PDF tails as observables of choice.  \\ \\

As we just saw, for large values of $\tilde{\delta}_\text{max}$, it is irrelevant if the largest jump is made in step one or two. However, for large values of $\tilde{x}_2=\tilde{\delta}_1+\tilde{\delta}_2$, it is relevant. Assuming the maximum of the IID random variables is the first one $\delta_\text{max}=\delta_1$ and also very large. Then the random walk is approximately $\tilde{x}_2 \approx (1+M_1) \delta_\text{max}$ where we used $\tilde{\delta}_1=\delta_\text{max}$ and $\tilde{\delta}_2\approx M_1 \delta_\text{max}$ for large $\delta_\text{max}$ as just explained. The crucial point is that not only the big jump $\tilde{\delta}_\text{max}=\delta_\text{max}$ but also the subsequent increment $\tilde{\delta}_2\approx M_1 \delta_\text{max}$ contributes to large values of $\tilde{x}_2$. On the other hand, assuming the maximum of the IID random variables is the second one $\delta_\text{max}=\delta_2$ and also very large. Then the random walk is approximately $\tilde{x}_2 \approx \delta_\text{max}$ where we used that $\tilde{\delta}_1$ can be neglected and $\tilde{\delta}_2\approx \delta_\text{max}$ for large $\delta_\text{max}$ as just explained. Both cases are summarized with the observation that the position of the random walker is approximately 
\begin{equation}\label{correlatedbigju}
\tilde{x}_2 \approx W_{2-b} \delta_\text{max}
\end{equation}
with the weights $W_0=1$ and $W_1=1+M_1$. Here, the value $b=1,2$ is the step number of maximum of the IID random variables, i.e. for $b=1$ we have $\delta_\text{max}=\delta_1$ and for $b=2$ we have $\delta_\text{max}=\delta_2$. Due to Eq.~(\ref{correlatedmaximu}), the relationship Eq.~(\ref{correlatedbigju}) can also be stated as
\begin{equation}\label{correlatedbigju2}
\tilde{x}_2 \approx W_{2-b} \tilde{\delta}_\text{max},
\end{equation}
which finally gives the relationship between $\tilde{x}_2$ and $\tilde{\delta}_\text{max}$ when both are large, see Fig.~\ref{fig:plot_05}. The scatter plot between $\tilde{x}_2$ and $\tilde{\delta}_\text{max}$ reveals the relevance of the step number $b$. The two possibilites of $\tilde{\delta}_\text{max}$ being the first or the second increment can be seen by the two lines in the scatter plot where each line has the slope $W_{2-b}$ as predicted by Eq.~(\ref{correlatedbigju2}) with $b=1$ or $2$. Furthermore, Eq.~(\ref{correlatedbigju2}) is the variable transform of the conditional BJP of Eq.~(\ref{cobjp2}). In Fig.~\ref{fig:fig_01-02_really_new} we presented the conditional distributions. Until now, we always conditioned on the step number $b$ of the maximum $\delta_\text{max}$ (below we also condition on the step number $b$ of $\tilde{\delta}_\text{max}$). In order to get the unconditional BJP Eq.~(\ref{unbjp1}) from the conditional BJP, we need to average the conditional distributions of $\tilde{x}_N$ and $\tilde{\delta}_\text{max}$ over all $b$ weighted with the probability of $b$. We will discuss this below.\\ \\

For the exponential memory kernel, it is straightforward to increase $N$ and get again similar statements. We just learned that for the first two increments $\tilde{\delta}_1$ and $\tilde{\delta}_2$ we find the relationships  Eq.~(\ref{correlatedmaximu}) and (\ref{correlatedbigju2}). Now we consider $N=3$. The third increment is $\tilde{\delta}_3=M_1 \tilde{\delta}_2 +\delta_3$ which has the same structure as the second increment $\tilde{\delta}_2=M_1\tilde{\delta}_1 +\delta_2$, i.e. it depends on the previous increment multiplied with $m$ plus a IID random variable. Therefore, above results Eq.~(\ref{correlatedmaximu}) and (\ref{correlatedbigju2}) are valid also for $N=3$. Generally for any finite $N$, we have the same structure $\tilde{\delta}_i = M_1 \tilde{\delta}_{i-1}+\delta_i$ so that we can claim generality of above results. For the algebraical memory kernel, it is  $\tilde{\delta}_i \neq M_1 \tilde{\delta}_{i-1}+\delta_i$. Nevertheless, we will derive the results of this section for both memory kernels and all $N$. In particular, that the large values of $\tilde{x}_2$ (and also $\tilde{x}_N$) depend on the step number $b$ of the big jump makes it necessary to describe two BJPs, namely unconditioned and conditioned on $b$.

\section{Unconditional Big Jump Principle}\label{sec:uncond}

We derive the unconditional BJP which is the relationship of the tail distributions between large values of the maximum $\tilde{\delta}_\text{max}$ and the sum $\tilde{x}_N$. In the next section we will consider the conditional distributions of these variables conditioned on the step number of $\tilde{\delta}_\text{max}$.

\subsection{Tail of the Maximum Distribution for $\tilde{\delta}_\text{max}$}\label{sec:tailmaxi}

\begin{figure}\begin{center}
\includegraphics[width=0.45\textwidth]{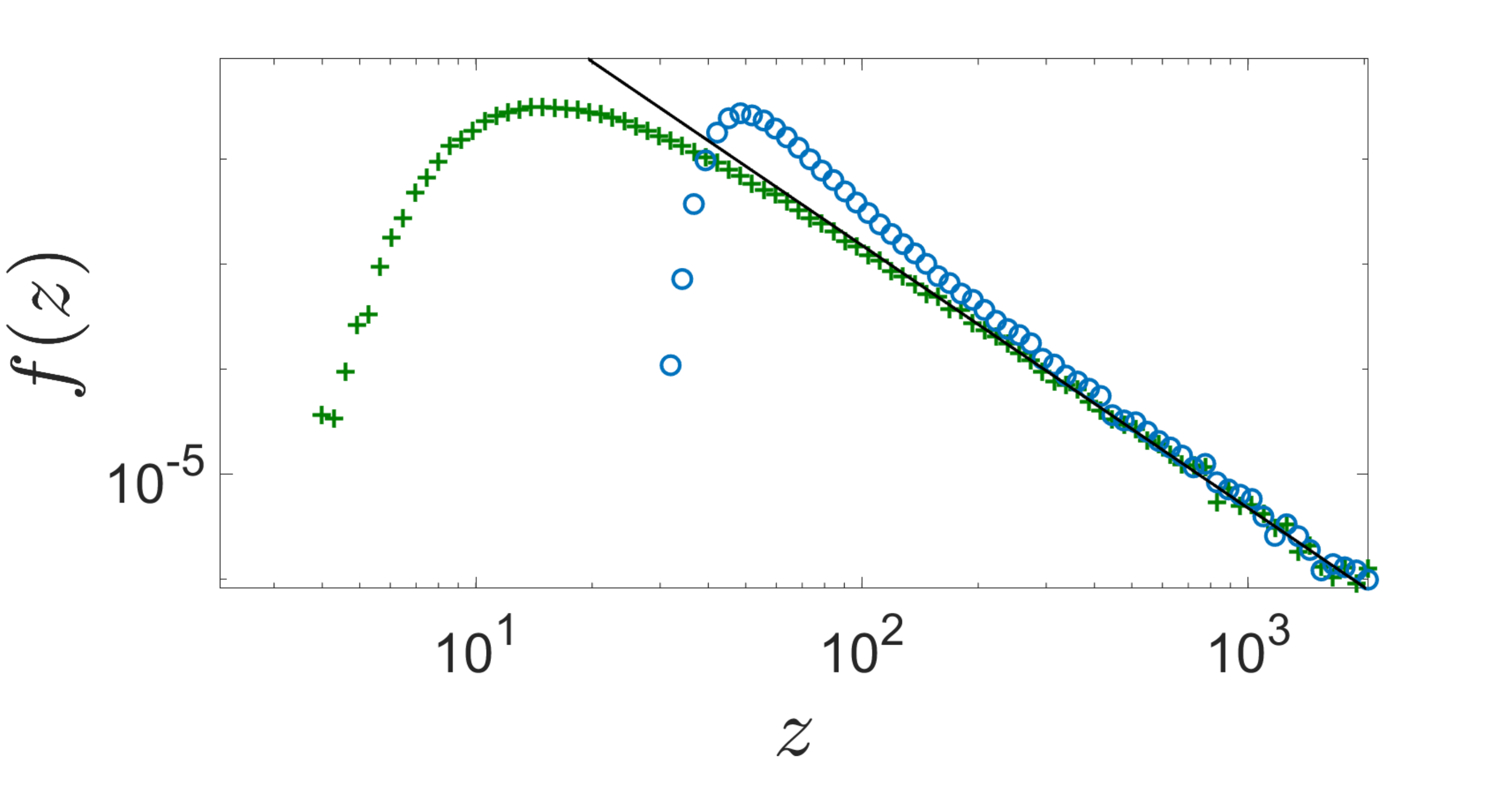}
\caption{Histograms for $\delta_\text{max}$ (green crosses) and $\tilde{\delta}_\text{max}$ (blue circles) compared with the theory of Eq.~(\ref{bjpmaximum}) (black line) where $N=100$. The $\delta_i$ follow the Pareto PDF with $\alpha=3/2$. The memory kernel is algebraical with the parameter $\beta=1/2$. The number of realizations is $10^5$. For large $z$ the two curves merge, an indication for the ``big jump principle'' between $\delta_\text{max}$ and $\tilde{\delta}_\text{max}$ at work. \label{fig:plot_06}}
\end{center}\end{figure}

\begin{figure}\begin{center}
\includegraphics[width=0.5\textwidth]{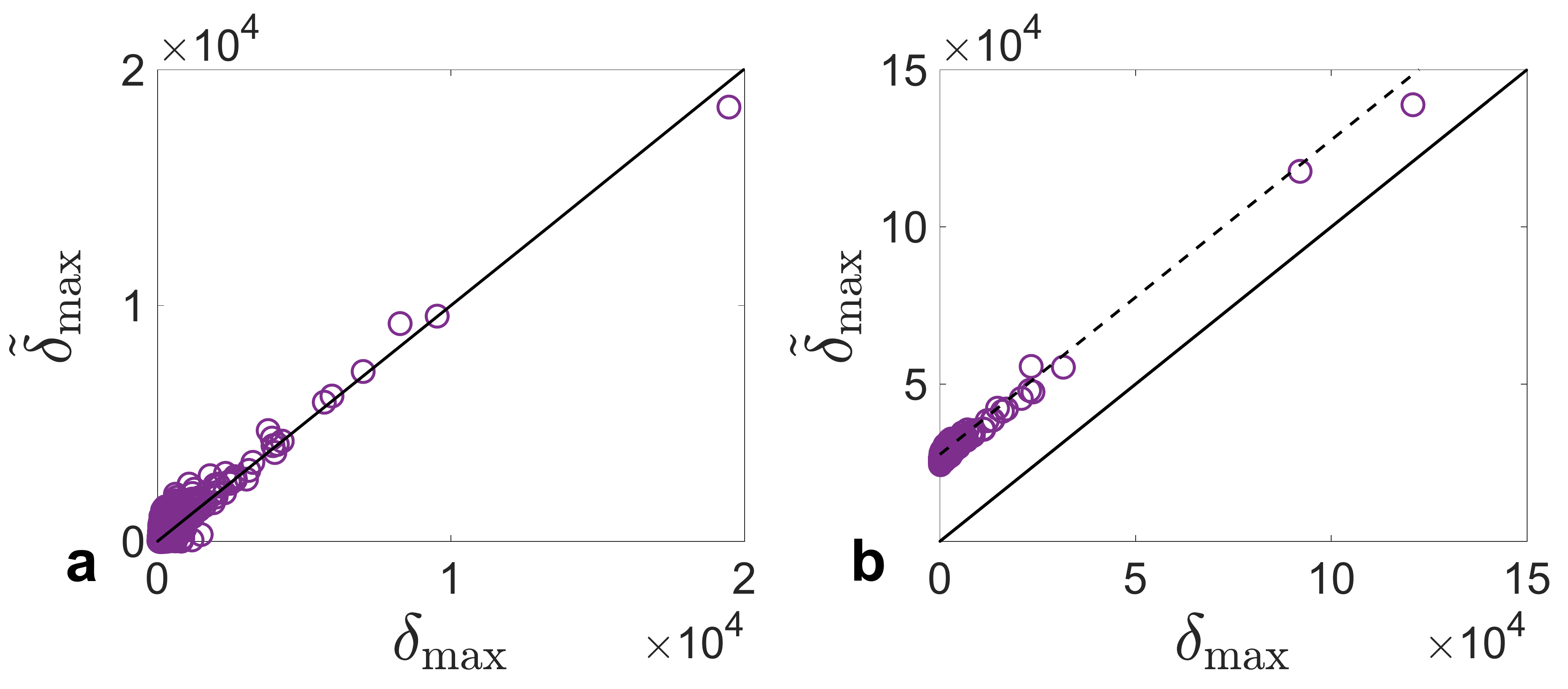}
\caption{\textbf{a} Scatter plot between $\delta_\text{max}$ and $\tilde{\delta}_\text{max}$ (purple circles) where $N=10^4$. The $\delta_i$ follow $L_{\alpha,0,1,0}(z)$. The memory kernel is algebraical with $\beta=0.01$. The number of realizations is $10^3$. A line with slope one is shown (black line). There is a linear relationship between $\delta_\text{max}$ and $\tilde{\delta}_\text{max}$ when both are large as predicted by Eq.~(\ref{bjpmaximum}). \textbf{b} The same plot as the left figure but with the $\delta_i$ following the Pareto PDF with $\alpha=3/2$. The big jump principle follows a linear line $\tilde{\delta}_b \approx \delta_b + \langle \delta_i \rangle \sum_{b=1}^N \left[\Phi(b) \sum_{j=1}^{b-1}M_{b-j} \right]$ (dashed line) as explained in the main text. \label{fig:plot_07-08}}
\end{center}\end{figure}

The unconditional cumulative distribution function (CDF) of the maximum is per definition
\begin{equation}\begin{split}
F_{\tilde{\delta}_\text{max}}(z;N)&=\text{Prob}(\tilde{\delta}_\text{max}\le z)\\
&=\text{Prob}(\tilde{\delta}_1 \le z ,\ldots , \tilde{\delta}_N \le z),
\end{split}\end{equation}
i.e. every increment $\tilde{\delta}_i$ is less or equal than the value $z$. In particular, here, it is irrelevant which increment is the maximum. Here and only in this subsection, we add $N$ in the argument of $F_{\tilde{\delta}_\text{max}}(z;N)$ to highlight the $N$-dependence. We always assume $N$ to be finite.\\ \\

We show in App.~\ref{sec:appmax} for the case with a power law PDF $f_{\delta_i}(z) \sim A z^{-1-\alpha}$ that the large $z$ behaviour of $F_{\tilde{\delta}_\text{max}}(z;N)$ is given by the product 
\begin{equation}\label{corrmaxcdf}
F_{\tilde{\delta}_\text{max}}(z;N) \sim F_{\tilde{\delta}_\text{max}}(z;N-1) F_{\delta_i}(z).
\end{equation}
For $N=2$ we have clearly a multiplication  of two CDFs just like the IID case. This was precisely the point in previous Sec.~\ref{sec:example}. Generally, the formula Eq.~(\ref{corrmaxcdf}), which is valid only for large $z$, is similar the IID case where the well-known maximum CDF $F_{\delta_\text{max}}(z;N)=[F_{\delta_i}(z)]^N$ describes the maximum of the $N$ IID random variables $\delta_i$. However, in the correlated case, only the first increment $\tilde{\delta}_1=\delta_1$ has the same statistical properties as those of the IID process. This gives the term $F_{\delta_i}(z)$ in Eq.~(\ref{corrmaxcdf}). The following $N-1$ increments $\tilde{\delta}_i$, $i\ge 2$, are correlated and therefore are governed by the term $F_{\tilde{\delta}_\text{max}}(z;N-1)$ instead of $F_{\delta_\text{max}}(z;N-1)=[F_{\delta_i}(z)]^{N-1}$. \\ \\

The maximum PDF is the derivative $f_{\tilde{\delta}_\text{max}}(z;N) = \mathrm{d}/\mathrm{d}z F_{\tilde{\delta}_\text{max}}(z;N) $ and therefore
\begin{equation}\label{induction}
f_{\tilde{\delta}_\text{max}}(z;N) \sim f_{\tilde{\delta}_\text{max}}(z;N-1)+ f_{\delta_i}(z).
\end{equation}
Here, we used that both CDFs $F_{\tilde{\delta}_\text{max}}(z;N-1)$ and $F_{\delta_i}(z)$ are about $1$ for large $z$. Thus, we can deduce (proof by induction) that the tail of the maximum increment PDF follows
\begin{equation}\label{tailmaxi25}
f_{\tilde{\delta}_\text{max}}(z;N) \sim N A z^{-1-\alpha} \sim N f_{\delta_i}(z)
\end{equation}
which is the same as the tail of $f_{\delta_\text{max}}(z;N)$ describing the maximum of the IID random variables. Summarized, we can say that
\begin{equation}\label{bjpmaximum}
\text{Prob}(\tilde{\delta}_\text{max}>z)\sim\text{Prob}(\delta_\text{max}>z)
\end{equation}
for large $z$, see Fig.~\ref{fig:plot_06}. This formula is the general statement of the exemplary results of Eq.~(\ref{correlatedmaximu}). \\ \\

In Fig.~\ref{fig:plot_07-08}, we present the scatter plot between $\tilde{\delta}_\text{max}$ and $\delta_\text{max}$ for the algebraical memory kernel for two different PDFs of $\delta_i$, namely the two-sided symmetric L\'evy PDF and the Pareto PDF. For the first example, we observe a straight line for large values $\tilde{\delta}_\text{max}\approx \delta_\text{max}$. But for the second example, we see some offset $\tilde{\delta}_\text{max}\approx \delta_\text{max}+C$. The reason for that is the correction to the BJP which we already found for the IID case, see Sec.~\ref{sec:random} and App.~\ref{sec:correction}. We can find the offset $C$ here by replacing the remaining random (which are not the maximum) variables by their mean, see details for the correlated random walk also in App.~\ref{sec:correction}. Let's say $\tilde{\delta}_\text{max}$ happens at b so that $\tilde{\delta}_\text{max}=\tilde{\delta}_b$. Per definition, the $b$-th increment is $\tilde{\delta}_b=\delta_b+\sum_{j=1}^{b-1} M_{b-j} \delta_j$. Assuming $\delta_b$ is large, we claim $\tilde{\delta}_b \stackrel{d}{\sim} \delta_b + \langle \delta_i \rangle \sum_{j=1}^{b-1} M_{b-j}$. This formula is conditioned on the step number $b$. Now we average over $b$. Thus, we get the unconditional relationship $\tilde{\delta}_\text{max} \stackrel{d}{\sim} \delta_\text{max}+C$ with the offset $C=\langle \delta_i \rangle \sum_{b=1}^N \left[\Phi(b) \sum_{j=1}^{b-1}M_{b-j} \right]$. The probability $\Phi(b)$ that $\tilde{\delta}_\text{max}$ appears at $b$ will be discussed below when we study the conditional BJP. In Fig.~\ref{fig:plot_07-08}, the offset $C$ is visible in the scatter plot for the Pareto case because small $\beta$ and large $N$ lead to a large $C$. If we would sample more data, we eventually would find even larger maxima so that the offset will be less visible in the scatter plot. But even for small $C$ we find that this correction describes well also other distributions of this work at hand. In App.~\ref{sec:correction}, we present the correction terms for the conditional and unconditional sum and maximum distributions.

\subsection{Tail of the Sum Distribution for $\tilde{x}_N$}
We already calculated the unconditional PDF of $\tilde{x}_N$ when the IID random variables $\delta_i$ follow the symmetric L\'evy PDF, see Eq.~(\ref{largenlimit1}). We also found the PDF of $\tilde{x}_N$ for the Pareto case when $N\to\infty$, see Eq.~(\ref{largenlimit2}). Hence, we are left with the Pareto case when $N$ is finite which we perform now. \\ \\

The unconditional PDF of the sum $\tilde{x}_N$ is the $N$-fold convolution
\begin{equation}
f_{\tilde{x}_N}(z) = (f_{W_{N-1}\delta_1}\ast \ldots  \ast f_{W_0\delta_N})^{(N)}(z).
\end{equation}
The effect of correlations enters in this equation through the factors $W_{N-1},\ldots , W_0$. The $2$-fold convolution is defined as $(f\ast g)^{(2)}(z)=\int_0^z f(x) g(z-x)\mathrm{d}x$ and higher orders subsequently. We calculate the Laplace transform $\hat{f}_{\tilde{x}_N}(s)=\int_0^\infty f_{\tilde{x}_N}(z) \text{exp}(-zs)\mathrm{d}z$. The $N$-fold convolution is in Laplace space the $N$-product
\begin{equation}
\hat{f}_{\tilde{x}_N}(s) = \prod_{i=1}^N\hat{f}_{\delta_i}(W_{N-i}s).
\end{equation}
The small $s$ behaviour of the Laplace transform for the increment Pareto PDF is $\hat{f}_{\delta_i}(s) \sim 1 + \alpha \Gamma(-\alpha)s^\alpha$ when $\alpha\in(0,1)$ and $\hat{f}_{\delta_i}(s) \sim 1-\langle \delta_i \rangle s + \alpha \Gamma(-\alpha)s^\alpha$ when $\alpha\in(1,2)$. Therefore, we find the large sum behaviour given by the small $s$ expansion $\hat{f}_{\tilde{x}_N}(s) \sim \alpha \Gamma(-\alpha) \tilde{\gamma}_Ns^\alpha$ for both $\alpha\in(0,1)$ and $\alpha\in(1,2)$ with the scaling factor $\tilde{\gamma}_N=\sum_{i=1}^N (W_{N-i})^\alpha$ (see Eq.~(\ref{scalingfa})). Inverse Laplace transform $s\to z$ gives the large z behaviour
\begin{equation}\label{tailx}
f_{\tilde{x}_N}(z) \sim  \tilde{\gamma}_N \alpha z^{-1-\alpha} .
\end{equation}
We finally have all results to state the unconditional BJP.

\subsection{Unconditional Big Jump Principle}

\begin{figure}\begin{center}
\includegraphics[width=0.45\textwidth]{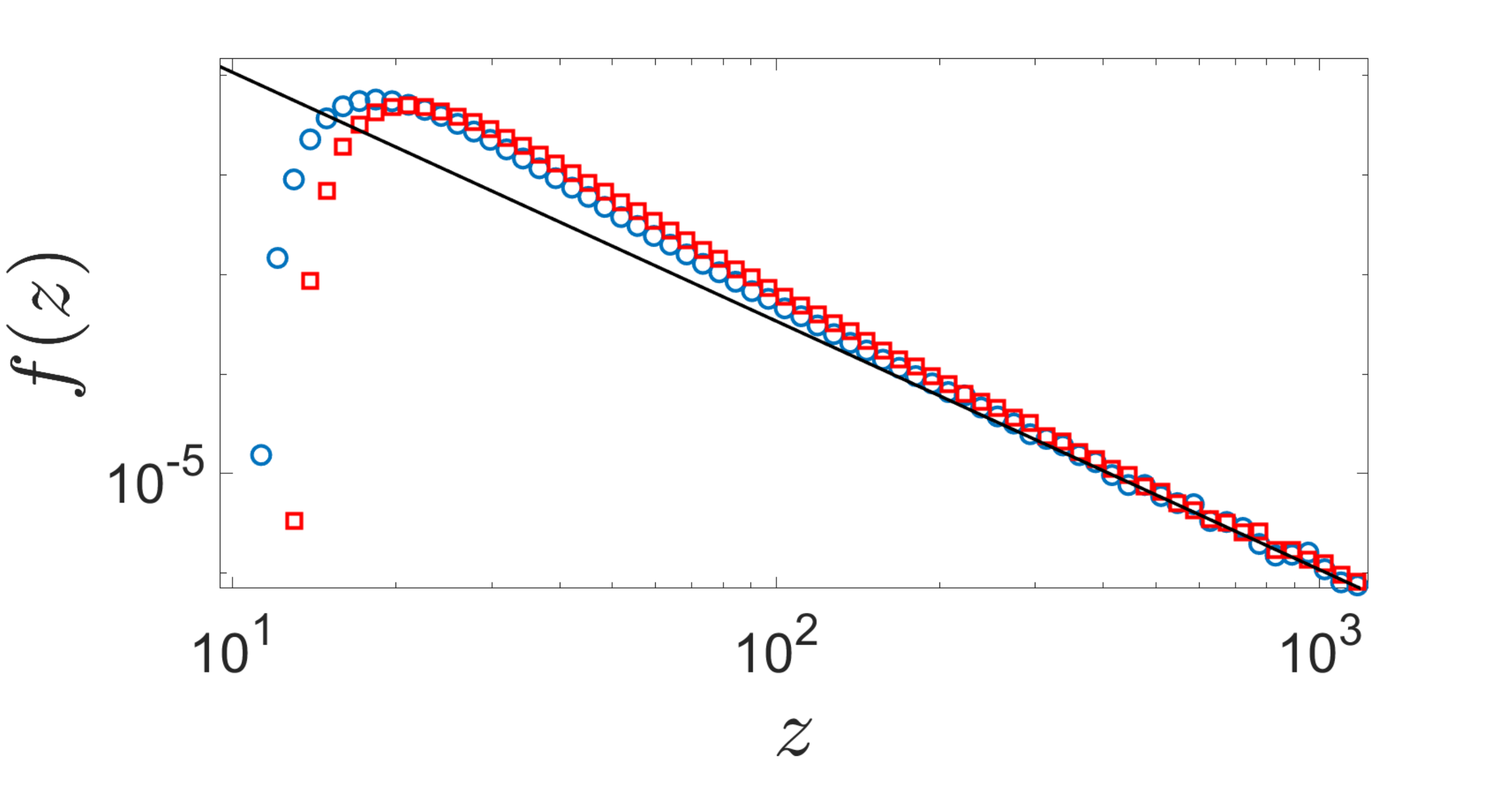}
\caption{Histogram for $(\tilde{\gamma}_N/N)^{1/\alpha}\tilde{\delta}_\text{max}$ (blue circles) with $(\tilde{\gamma}_N/N)^{1/\alpha}\approx 2.745$ and $\tilde{x}_N$ (red squares) compared with the unconditional BJP of Eq.~(\ref{bjpweighted}) (black line) where $N=5$. The $\delta_i$ follow the Pareto PDF with $\alpha=3/2$. The memory kernel is exponential with the parameter $m=0.9$. The number of realizations is $10^6$.  \label{fig:plot_09}}
\end{center}\end{figure}

We compare the tail of the maximum distribution of Eq.~(\ref{tailmaxi25}) with the tail of the sum distributions of Eq.~(\ref{largenlimit1}) and (\ref{tailx}). We find the unconditional BJP
\begin{equation}\label{bjpweighted}
\text{Prob}(\tilde{x}_N>z) \sim \frac{\tilde{\gamma}_N}{N} \text{Prob}\left(\tilde{\delta}_\text{max}>z\right)
\end{equation}
for large $z$ and any $N$. We presented it earlier in Eq.~(\ref{unbjp1}). This formula implies the random variable transform
\begin{equation}\label{bjpweighted45}
\tilde{x}_N \stackrel{d}{\sim} \left( \frac{\tilde{\gamma}_N}{N} \right)^{1/\alpha} \tilde{\delta}_\text{max}
\end{equation}
for large values because $f_{ \left(\tilde{\gamma}_N/N \right)^{1/\alpha} \tilde{\delta}_\text{max}}(z)\sim(\tilde{\gamma}_N/N) f_{  \tilde{\delta}_\text{max}}(z)$. In Fig.~\ref{fig:plot_09} we compare the theory of Eq.~(\ref{bjpweighted}) with Monte-Carlo simulations and find perfect agreement.

\section{Conditional Big Jump Principle}\label{sec:cond}

\begin{figure}\begin{center}
\includegraphics[width=0.5\textwidth]{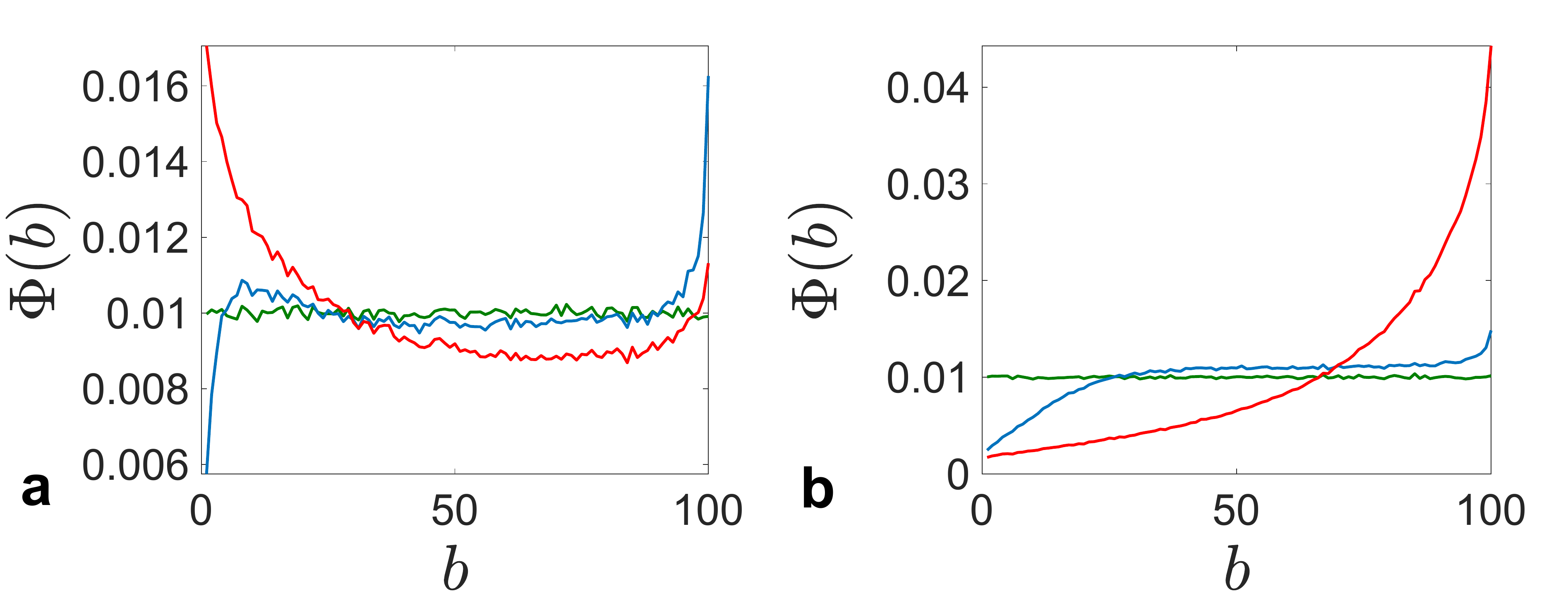}
\caption{\textbf{a} Probability $\Phi(b)$ that the big jump $\tilde{\delta}_\text{max}$ occurs at $b$ with $N=100$ for no memory (green), exponential memory wiht $m=0.9$ (blue) and algebraical memory with $\beta=0.5$ (red). The $\delta_i$ follow the two-sided L\'evy PDF with $\alpha=1.5$. The number of realizations is $10^6$. \textbf{b} The same figure as the left figure but the IID random variables $\delta_i$ follow the Pareto PDF with $\alpha=1.5$.  \label{fig:plot_10-11}}
\end{center}\end{figure}

\begin{figure}\begin{center}
\includegraphics[width=0.43\textwidth]{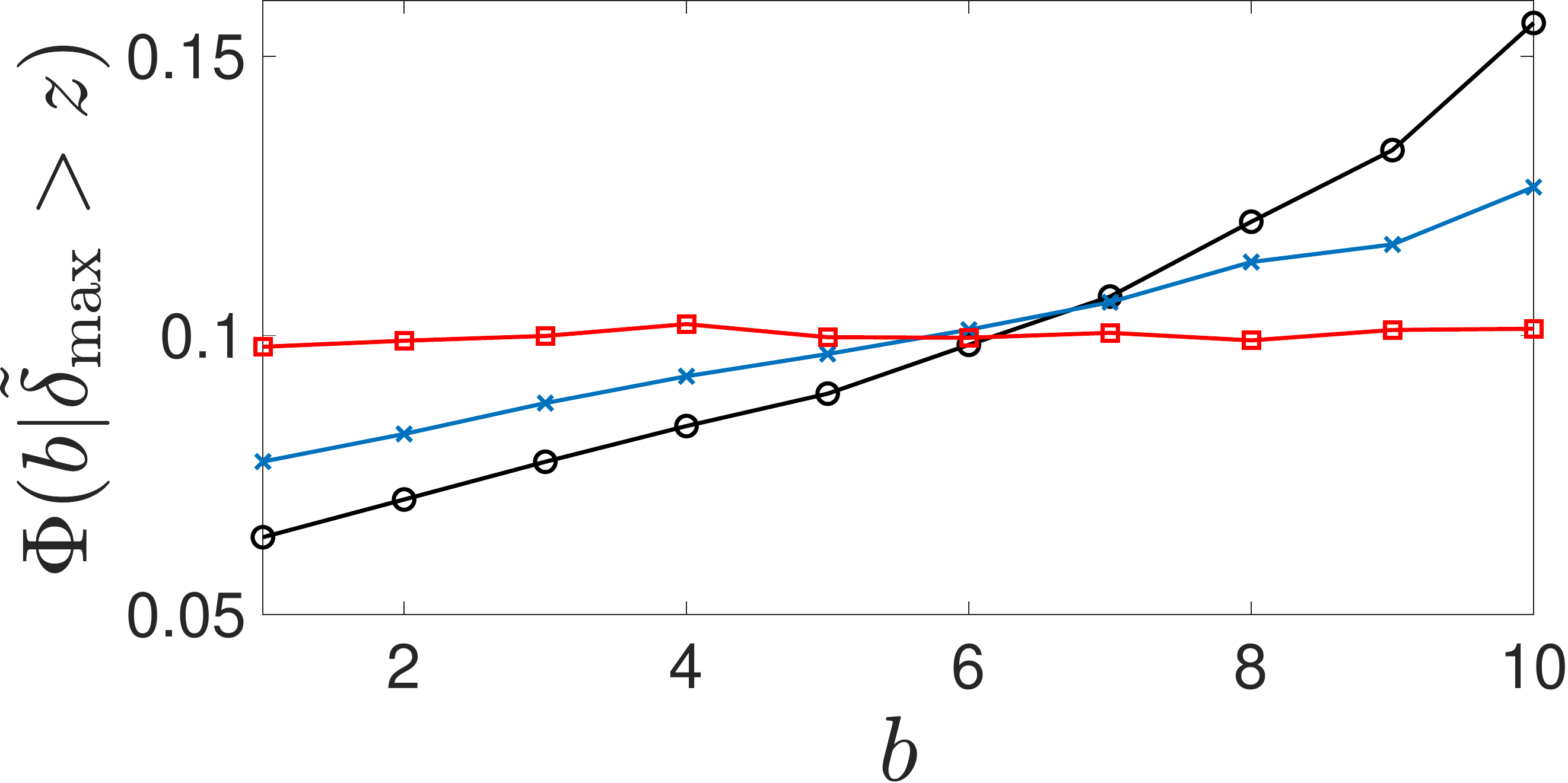}
\caption{Conditional probability $\Phi(b|\tilde{\delta}_\text{max}>z)$ for $z=1$ (black), $z=100$ (blue) and $z=10^5$ (red). The $\delta_i$ follow the Pareto PDF with $\alpha=0.5$. The memory kernel is algebraical with $\beta=0.5$. We used $N=10$ and $10^5$ realizations. \label{fig:plot_11a}}
\end{center}\end{figure}

\begin{figure}[t]\begin{center}
\includegraphics[width=0.43\textwidth]{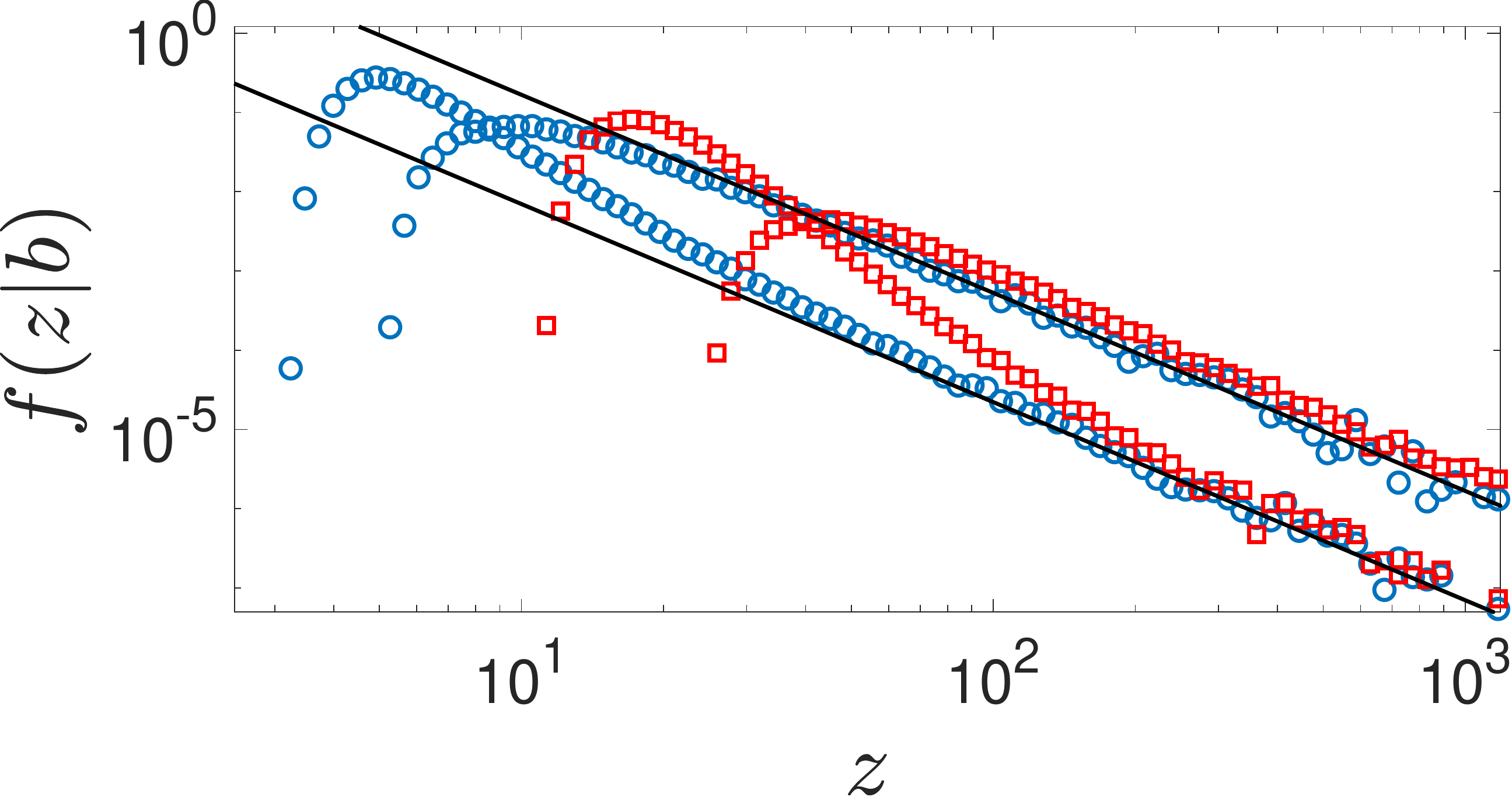}
\caption{Histograms for $\tilde{\delta}_\text{max}$ (blue circles) and $\tilde{x}_N/W_{N-b}$ (red squares) both conditioned on the step number of the big jump $\tilde{\delta}_\text{max}$ compared with the conditional BJP of Eq.~(\ref{bjpcond}) (black lines) where $N=5$, i.e. the black lines are $\alpha z^{-1-\alpha}/\Phi(b)$. The three curves matching in the upper line represent $b=1$ and the three curves matching in the lower line represent $b=5$. As explained in the main text, we estimated $\Phi(1)\approx 0.0284$ and $\Phi(5) \approx 0.6795$. The $\delta_i$ follow the Pareto PDF with $\alpha=1.5$. The memory kernel is algebraical with the parameter $\beta=0.5$. The number of realizations is $10^6$.  \label{fig:plot_12}}
\end{center}\end{figure}

\begin{figure}[t]\begin{center}
\includegraphics[width=0.43\textwidth]{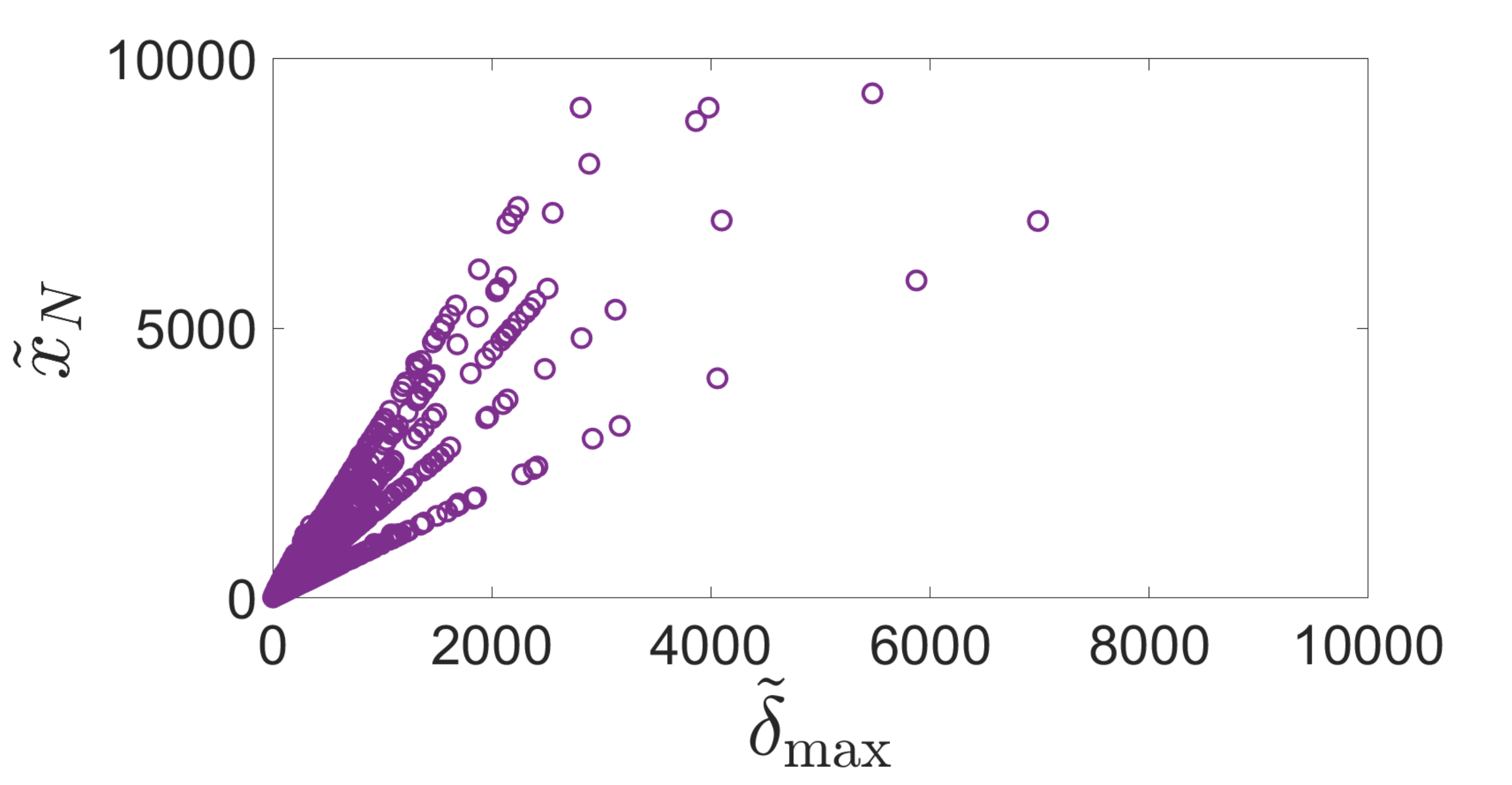}
\caption{Scatter plot between $\tilde{\delta}_\text{max}$ and $\tilde{x}_N$ (purple circles) where $N=5$. The $\delta_i$ follow the Pareto PDF with $\alpha=3/2$. The memory kernel is algebraical with $\beta=1/2$. The number of realizations is $10^6$. The five lines have the slopes $W_0,W_1,\ldots,W_4$. The five linear relationships between $\tilde{\delta}_\text{max}$ and $\tilde{x}_N$ when both are large are predicted in Eq.~(\ref{scatterlines}) and indicate at which step $\tilde{\delta}_\text{max}$ occurred.  \label{fig:plot_13}}
\end{center}\end{figure}

In contrast to the IID case, the step number when the big jump happens plays an important role. For example, when we observe the scatter plot between $\tilde{x}_N$ and $\tilde{\delta}_\text{max}$ we don't observe a single line with the slope given by the prefactor $(\tilde{\gamma}_N/N)^{1/\alpha} $ of Eq.~(\ref{bjpweighted45}), see Fig.~\ref{fig:plot_05}. The reason is that Eq.~(\ref{bjpweighted}) refers to the unconditional distributions and the scatter plot depends on the time steps when the big jump occurs, i.e. we are dealing with conditional distributions.  \\ \\

Generally, the unconditional PDFs are
\begin{equation}\begin{split}\label{condi}
f_{\tilde{x}_N}(z) &= \sum_{b=1}^N  \Phi(b) f_{\tilde{x}_N|b}(z|b), \\
f_{\tilde{\delta}_\text{max}}(z) &= \sum_{b=1}^N \Phi(b) f_{\tilde{\delta}_\text{max}|b}(z|b) .
\end{split}\end{equation}
The conditional PDFs $f_{\tilde{x}_N|b}(z|b)$ and $f_{\tilde{\delta}_\text{max}|b}(z|b)$ depend on the time step $b=1,\ldots,N$ when the maximum $\tilde{\delta}_\text{max}$ occurs. $\Phi(b)$ is the probability that $\tilde{\delta}_\text{max}$ happens at $b$. This probability can be numerically obtained, see Fig.~\ref{fig:plot_10-11}. There we compare $\Phi(b)$ of both memory kernels with the probability that the IID big jump $\delta_\text{max}$ happened at $b$. The latter is obviously $\Phi(b)=1/N$. Before we continue, we have to discuss a subtle issue  regarding $\Phi(b)$. We presented previously the relationship between large values of $\tilde{\delta}_\text{max}$ and of $\delta_\text{max}$, see Eq.~(\ref{bjpmaximum}). Both large values are equally distributed. How can this be when the probability $\Phi(b)$ of $\tilde{\delta}_\text{max}$ being at $b$ is not $1/N$? To answer this question, we have to consider the conditional probability $\Phi(b|\tilde{\delta}_\text{max}>z)$ that $\tilde{\delta}_\text{max}$ larger than some value $z$ . For large values $z$, we find $\Phi(b|\tilde{\delta}_\text{max}>z) \sim 1/N$, see Fig.~\ref{fig:plot_11a}. Therefore, we confirm our findings that only the large values of $\tilde{\delta}_\text{max}$ and of $\delta_\text{max}$ are statistically equal.\\ \\

From the large $z$ behaviour of the unconditional PDFs $f_{\tilde{x}_N}(z)$ of Eq.~(\ref{largenlimit1}) and (\ref{tailx}) and $f_{\tilde{\delta}_\text{max}}(z)$ of Eq.~(\ref{tailmaxi25}), we can conclude the large $z$ behaviour of the conditional PDFs from Eq.~(\ref{condi}). We get
\begin{equation}\begin{split}\label{condi4}
f_{\tilde{x}_N|b}(z|b) &\sim \frac{(W_{N-b})^\alpha}{\Phi(b)} Az^{-1-\alpha}, \\
f_{\tilde{\delta}_\text{max}|b}(z|b) &\sim \frac{1}{\Phi(b)} Az^{-1-\alpha},
\end{split}\end{equation}
see Fig.~\ref{fig:plot_12}. On a side note, if we would condition on the occurrence of $\delta_\text{max}$ and not $\tilde{\delta}_\text{max}$ both conditional PDFs behave as $f_{\tilde{x}_N|b}(z|b) \sim (W_{N-b})^\alpha N Az^{-1-\alpha}$ and $f_{\tilde{\delta}_\text{max}|b}(z|b) \sim N Az^{-1-\alpha}$ because here $\Phi(b)=1/N$. In particular, the conditional maximum PDF has the same large $z$ behaviour as the unconditional one, but this is only true if we condition on the IID big jump $\delta_\text{max}$. In Sec.~\ref{sec:example}, we considered the condition on $\delta_\text{max}$.  \\ \\

When we compare the large $z$ behaviour of the conditional PDFs Eq.~(\ref{condi4}), we find the conditional BJP 
\begin{equation}\label{bjpcond}
\text{Prob}(\tilde{x}_N>z|b) \sim  (W_{N-b})^{\alpha} \text{Prob}\left(\tilde{\delta}_\text{max}>z|b\right)
\end{equation}
conditioned on the occurrence of the big jump $\tilde{\delta}_\text{max}$ at $b$ for large $z$ and any $N$, see Fig.~\ref{fig:plot_12}. We presented this result earlier in Eq.~(\ref{cobjp2}). The conditional BJP is governed by the random variable transform
\begin{equation}\label{scatterlines}
\tilde{x}_N \stackrel{d}{\sim} W_{N-b} \tilde{\delta}_\text{max}
\end{equation}
conditioned on the step number $b$. If the biggest jump was the last one, $b=N$, the correlations don't play a role, while the effect is largest if $b=1$. This equation is the generalisation of the exemplary result of Eq.~(\ref{correlatedbigju2}). It can be visualised in a scatter plot between $\tilde{x}_N$ and $ \tilde{\delta}_\text{max}$. For large values of the random variables, one sees $N$ lines with each the slope $W_{N-b}$ depending on $b$. In Fig.~\ref{fig:plot_01-04} it was presented for $N=2$, in Fig.~\ref{fig:plot_13} we present $N=5$.\\ \\

\subsection{Alternative Argumentation}

\begin{figure}\begin{center}
\includegraphics[width=0.43\textwidth]{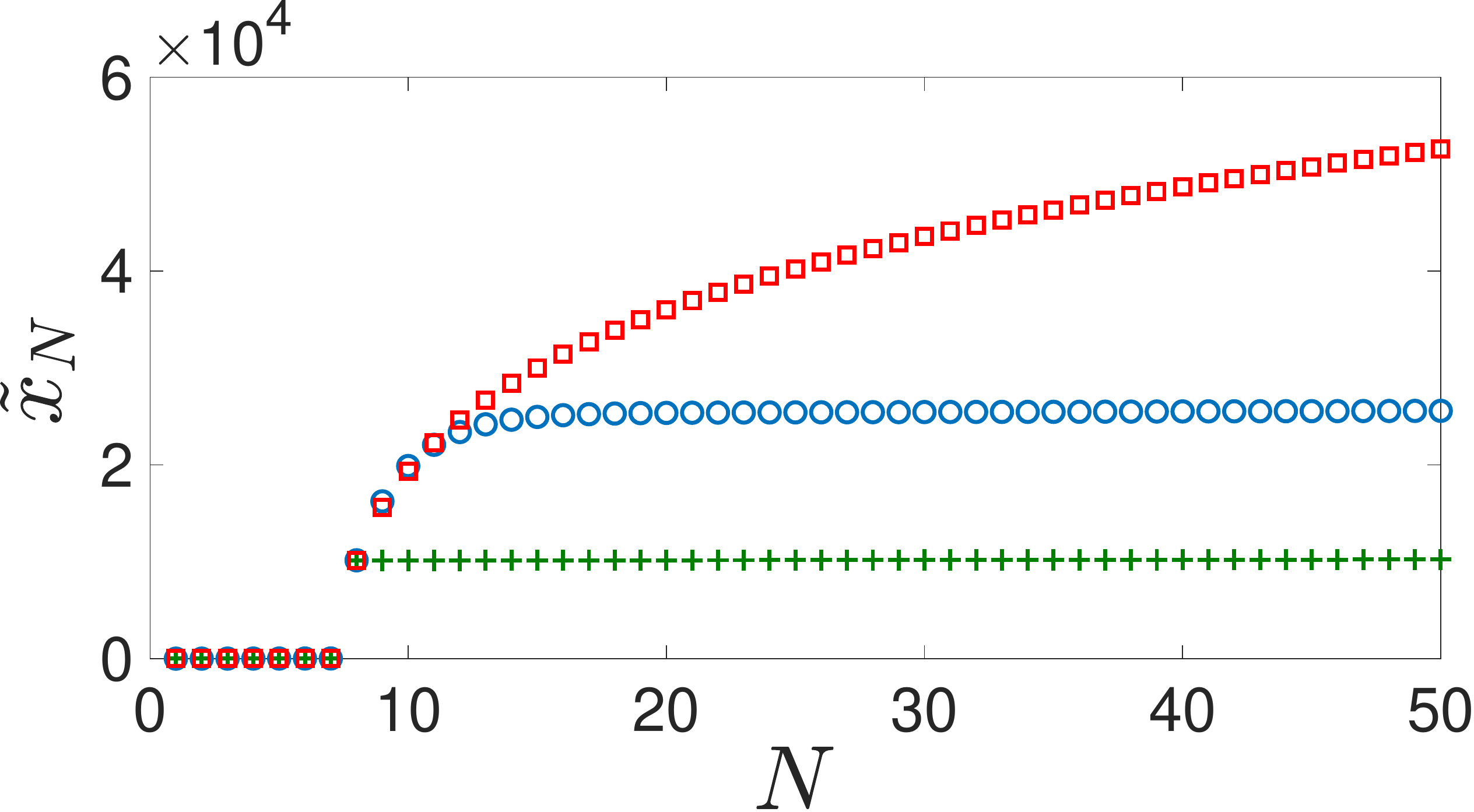}
\caption{The random walk $\tilde{x}_N$ with no correlations (green crosses), exponential memory kernel with $m=0.6$ (blue circles) and algebraical memory kernel with $\beta=0.9$ (red squares). One realization of $(\delta_1,\ldots,\delta_N)$ was used where the $\delta_i$ follow the Pareto PDF with $\alpha=3/2$. The big jump $\tilde{\delta}_\text{max} \approx 10128$ happend at $b=7$. The graph clearly shows the amplified influence of the big jump on the random walk due to the correlations. \label{fig:plot_14}}
\end{center}\end{figure}

We present an alternative argumentation to find the conditional BJP. The interesting point here is that we can use the IID BJP and find easily the conditional BJP without long calculations. In some sense, this alternative discussion here is a generalisation of Sec.~\ref{sec:example}. Most results in this subsection were shown previously. However, one particular point becomes clearer now, namely the behaviour of the subsequent increments after the big jump.\\ \\

We rearrange the definition of the sum $\tilde{x}_N=\sum_{i=1}^N\left(\sum_{j=1}^i M_{i-j}\delta_j\right)$ according to
\begin{equation}\label{rearrangem5}
\tilde{x}_N = \sum_{l=0}^{N-1} M_l \sum_{k=1}^{N-l} \delta_k=\sum_{l=0}^{N-1} M_l x_{N-l},
\end{equation}
i.e. the correlated sum $\tilde{x}_N$ is written in terms of a weighted sum of the uncorrelated sum $x_N$. Let's assume that the maximum of the IID random variables $\delta_\text{max}$ happens at step number $b$ and is also large. This is the condition on the appearance of $\delta_\text{max}$ and not $\tilde{\delta}_\text{max}$. This implies $x_{N-l} \stackrel{d}{\sim} \delta_\text{max}$ for $N-l\ge b$ which is the IID BJP of Eq.~(\ref{bjpintro}). Thus, Eq.~(\ref{rearrangem5}) becomes
\begin{equation}\label{reduction}
\tilde{x}_N \stackrel{d}{\sim}  \sum_{l=0}^{N-b} M_l\delta_\text{max}
= W_{N-b}\delta_\text{max}\end{equation}
where the terms with $N-l<b$ can be neglected. This is Eq.~(\ref{scatterlines}).\\ \\

Eq.~(\ref{reduction}) has an important implication, namely all increments with $i\ge b$ are statistically described by a single term
\begin{equation}\label{reduction2}
\tilde{\delta}_i \stackrel{d}{\sim} M_{i-b} \delta_\text{max}.
\end{equation}
This is a relation between the $i$-th increment $\tilde{\delta}_i$ with $i=b,\ldots,N$ and the maximum of the noise increments $ \delta_\text{max}=\delta_b$. Eq.~(\ref{reduction2}) can be seen by comparison of the prefactors between Eq.~(\ref{reduction}) and the definition $\tilde{\delta}_i = \sum_{j=0}^i M_{i-j}\delta_j$. Thus, the sum $ \sum_{j=0}^i M_{i-j}\delta_j$ is solely described by the single term $M_{b-i}\delta_\text{max}$ which is influenced by the big jump directly and all other summands can be neglected. Based on Eq.~(\ref{reduction2}), we find three key points under the assumption of large $\delta_\text{max}$:
\begin{itemize}
\item[1)] The big jump is statistically the same as the maximum of the IID random variables $\tilde{\delta}_\text{max} \stackrel{d}{\sim} \delta_\text{max}$ when both are large.
\item[2)] The increments following the big jump can never be neglected when summing up to $\tilde{x}_N$.
\item[3)] The correlated sum conditioned on $b$ is statistically $\tilde{x}_N \stackrel{d}{\sim} W_{N-b} \tilde{\delta}_\text{max}$ when both sides are large.
\end{itemize}
In particular, the third point is the conditional BJP of Eq.~(\ref{bjpcond}).\\ \\

We have seen how the interplay between big jump and correlations lead to the conditional BJP. Let's conclude this observation with another implication. Take the example of the IID random variables following the Pareto PDF $f_{\delta_i}(z)$. The non-negligible subsequent increments after the big jump lead to an amplification of the sum $\tilde{x}_N$ compared to $x_N$. We demonstrate this in Fig.~\ref{fig:plot_14} for the two memory kernels. The earlier the big jump happened the bigger is the influence of the big jump on the random walk due to the correlations. In other words, once a very large maximum occurred we can predict the future since the future depends only on the big jump (from the past) and the deterministic value of the correlation.

\subsection{Large $\bm{N}$ behaviour in the Scatter Plot}\label{sec:large}

\begin{figure}\begin{center}
\includegraphics[width=0.23\textwidth]{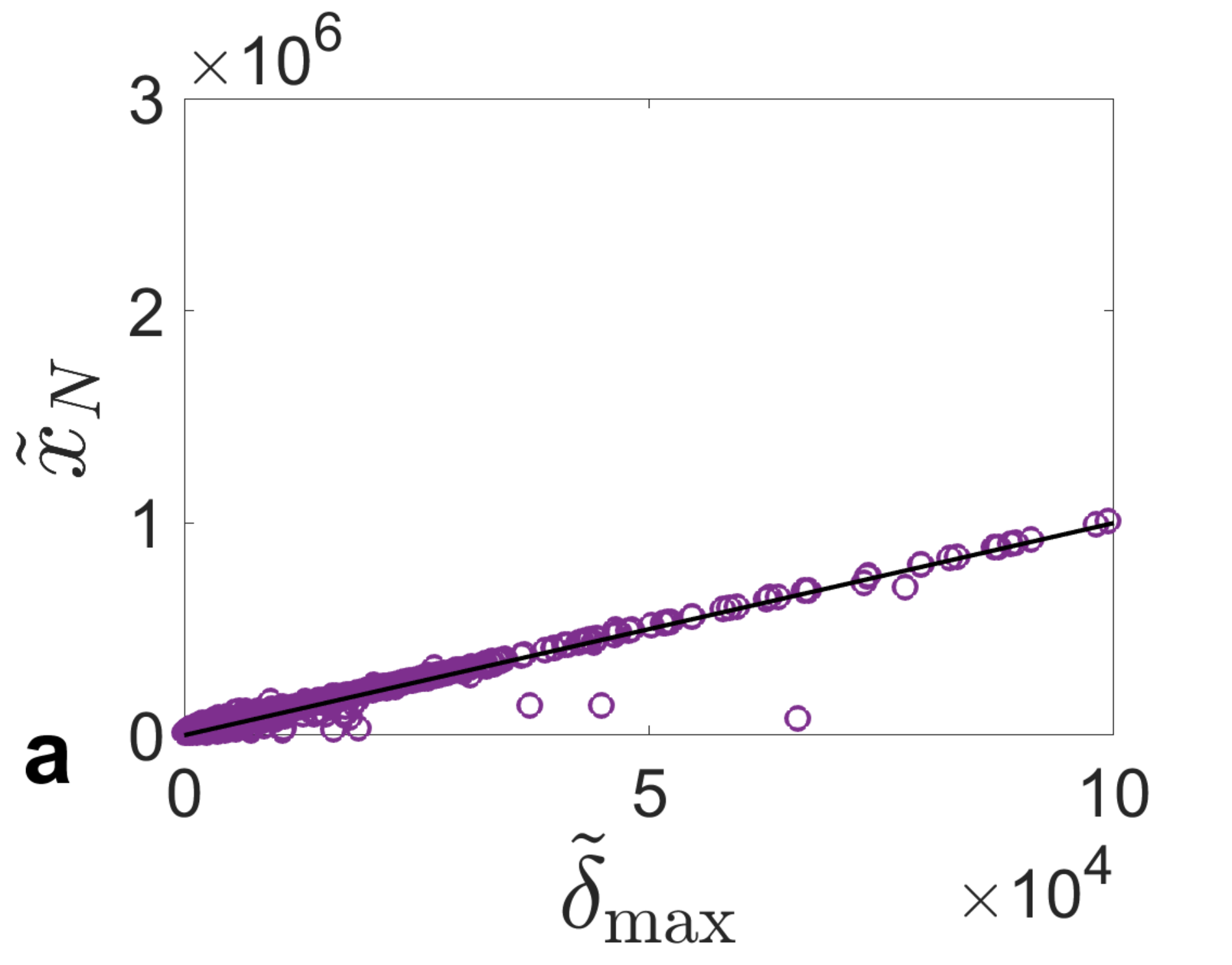}
\includegraphics[width=0.23\textwidth]{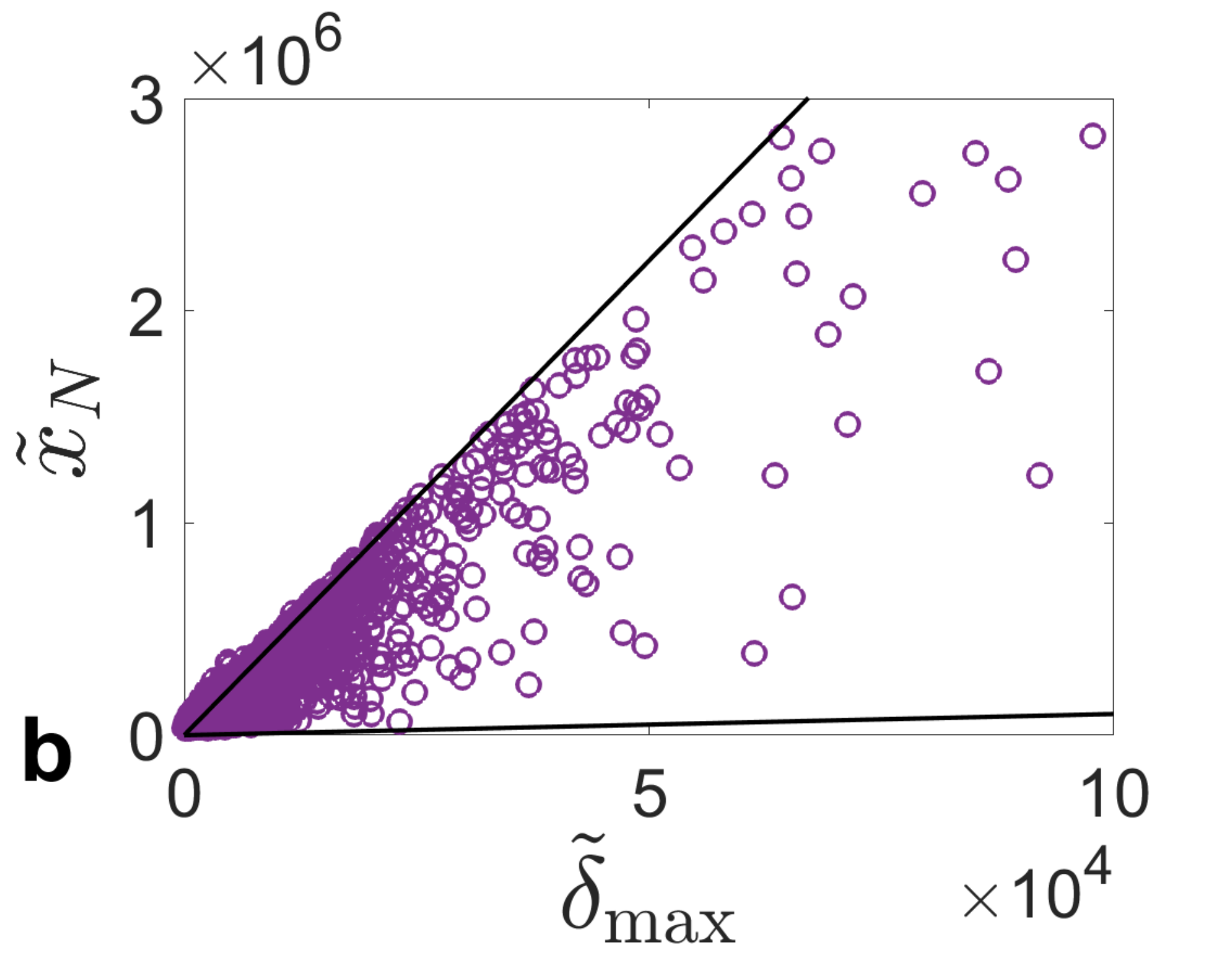}
\caption{\textbf{a} Scatter plot between $\tilde{\delta}_\text{max}$ and $\tilde{x}_N$ (purple circles) where $N=500$. The $\delta_i$ follow the Pareto PDF with $\alpha=1.5$. The memory kernel is exponential with $m=0.9$. The number of realizations is $10^6$. The relatively large number $N$ makes if statistically difficult to observe the predicted $N$ lines with slopes $W_0,\ldots,W_{N-1}$. Instead, most large values are located on the line with the slope $1/(1-m)$ (black line), see Eq.~(\ref{kernellimit}). \textbf{b} The same parameters as the left figure but the memory kernel is algebraical with $\beta=1/2$. We observe a triangle bordered by the two lines with slope $W_0=1$ and $N^{1-\beta}/(1-\beta)$ (two black lines), see Eq.~(\ref{kernellimit2}).
  \label{fig:plot_15-16}}
\end{center}\end{figure}

In Fig.~\ref{fig:plot_15-16}, we present the scatter plot between $\tilde{\delta}_\text{max}$ and $\tilde{x}_N$ for the large value of $N=500$. We present the exponential and algebraical memory kernel case. For both cases, we don't see $N$ lines as in the scatter plots of Fig.~\ref{fig:plot_01-04} and \ref{fig:plot_13} where we present small $N$. The reason is simply that we don't have enough realizations which are necessary to sample the rare events. Furthermore, we observe two different behaviours between the exponential and algebraical memory case. While exponential case yields a straight line, the algebraical case looks like a triangle. This is because the weight $W_{N-b}$ for large $N$ either converges (exponential memory) or not (algebraical memory). Here, it is important that we don't fix $b$ as previously in Eq.~(\ref{anotherlarge}) and (\ref{weightalg}). We explain now both cases in detail. \\ \\

For the exponential memory, we still can use the large $N$ behaviour of Eq.~(\ref{anotherlarge}), namely $W_{N-b}\sim 1/(1-m)$, even when we don't fix $b$. The latter is important. For large $N$ only values $b \approx N$ yield $W_{N-b}\not\sim 1/(1-m)$. But this amount of $W_{N-b}$ is negligible to those $W_{N-b}$ which behave as $1/(1-m)$. So we get the conditional BJP of Eq.~(\ref{bjpcond}) for large $N$ as
\begin{equation}\label{kernellimitfully}
\text{Prob}(\tilde{x}_N>z|b) \sim \frac{1}{(1-m)^\alpha} \text{Prob}\left(\tilde{\delta}_\text{max}>z|b\right)
\end{equation}
for large $z$. The prefactor in the right hand side doesn't depend on $b$. So this is also the unconditional BJP for large $N$ which we can show directly. When we take the large $N$ limit of the prefactor $\tilde{\gamma}_N/N$ in the unconditional BJP Eq.~(\ref{bjpweighted}), we find again Eq.~(\ref{kernellimitfully}) but now not conditioned on $b$. The variable transform of Eq.~(\ref{kernellimitfully}) is
\begin{equation}\label{kernellimit}
\tilde{x}_N \stackrel{d}{\sim} \frac{1}{1-m}\tilde{\delta}_\text{max}
\end{equation}
under the assumption of large $N$. The prefactor $1/(1-m)$ is the slope we observe in the scatter plot of  Fig.~\ref{fig:plot_15-16}. \\ \\

In contrast, for the algebraical memory kernel, the weight $W_{N-b}$ for large $N$ but not fixed $b$ cannot be replaced by $N^{1-\beta}/(1-\beta)$. The latter assumes fixed $b$, see Eq.~(\ref{weightalg}). For large $N$, there is a non-negligible amount of $W_{N-b}$ which is not similar to this limit. And this is what we observe in the scatter plot. It is statistically difficult to observe $N$ lines because it would require a tremendous amount of realizations of $\tilde{x}_N$. What one observe instead for a large but finite amount of realizations is a triangle structure confined by the lowest and largest slope
\begin{equation}\label{kernellimit2}
\tilde{\delta}_\text{max} \le \tilde{x}_N \le \frac{N^{1-\beta}}{1-\beta},
\end{equation}
see Fig.~\ref{fig:plot_15-16}. The upper bound diverges, so when $N\to\infty$ it is not really useful. One should hope it works for finite but large $N$. Of course, for a large enough number of realizations one would finally observe $N$ lines but this is practically difficult. In this sense the case of power law memory kernel is vastly different from the exponential case.

\section{Summary}\label{sec:outlook}

We studied the relationship between the maximum $\tilde{\delta}_\text{max}$ and the correlated sum $\tilde{x}_N$ in the limit when both are large. In contrast to the well-studied IID case, here the step number $b=1,\ldots,N$ of the big jump $\tilde{\delta}_\text{max}$ is relevant. We found three key observations under the assumption that the maximum of the IID random variables $\delta_\text{max}$ is large: 1) The big jump is statistically the same as the maximum of the IID random variables $\tilde{\delta}_\text{max}\stackrel{d}{\sim} \delta_\text{max}$ provided both are large, 2) the increments after the big jump are statistically described by $\tilde{\delta}_i \stackrel{d}{\sim} M_{i-b} \delta_\text{max}$ with $i\ge b$, see Eq.~(\ref{reduction2}), and 3) large values of the correlated sum conditioned on $b$ are statistically the same as the weighted big jump $\tilde{x}_N \stackrel{d}{\sim} W_{N-b}\tilde{\delta}_\text{max}$. In particular, the increments after the big jump contribute to the tail statistics of $\tilde{x}_N$. We described these mechanisms and, in particular, the dependence of the step number of the big jump on $\tilde{x}_N$ with two BJPs. These are the unconditional BJP of Eq.~(\ref{bjpweighted}) and the conditional BJP of Eq.~(\ref{bjpcond}). Beyond that, we believe that the results of this work are suitable for a generalisation when studying other correlated random walks. For example, the well-known time series analysis models ARMA and ARFIMA processes \cite{samoradnitsky2017stable,
burnecki2010fractional,
burnecki2017identification} can be written as a weighted sum of IID random variables. Furthermore, the studied techniques can be used in the evaluation of rare events in time series data where heavy-tails and correlations are present, e.g. transport of contamination in disordered media \cite{de2013flow,
berkowitz2010anomalous} and precipitation \cite{papalexiou2013battle}. \\ \\

\section*{Acknowledgment}

M.H. is funded by the Deutsche Forschungsgemeinschaft (DFG, German Research Foundation) – 436344834. E.B. acknowledges the
Israel Science Foundations Grant No. 1898/17. The authors thank H. Kantz and W. Wang for their helpful discussion and comments.

\begin{appendix}

\counterwithin{figure}{section}

\section{Correction term for the BJPs}\label{sec:correction}

\subsection{IID random variables}

We assume that the IID random variables $\delta_i$ follow the Pareto PDF with $\alpha>1$, i.e. the mean $\langle \delta_i \rangle$ exists. Clearly for the Pareto PDF all random variables are positive and hence $x_N>\delta_\text{max}$. Here, we find a correction to the IID BJP $x_N \stackrel{d}{\sim} \delta_\text{max}$ with the following idea.\\ \\

\begin{figure}\begin{center}
\includegraphics[width=0.43\textwidth]{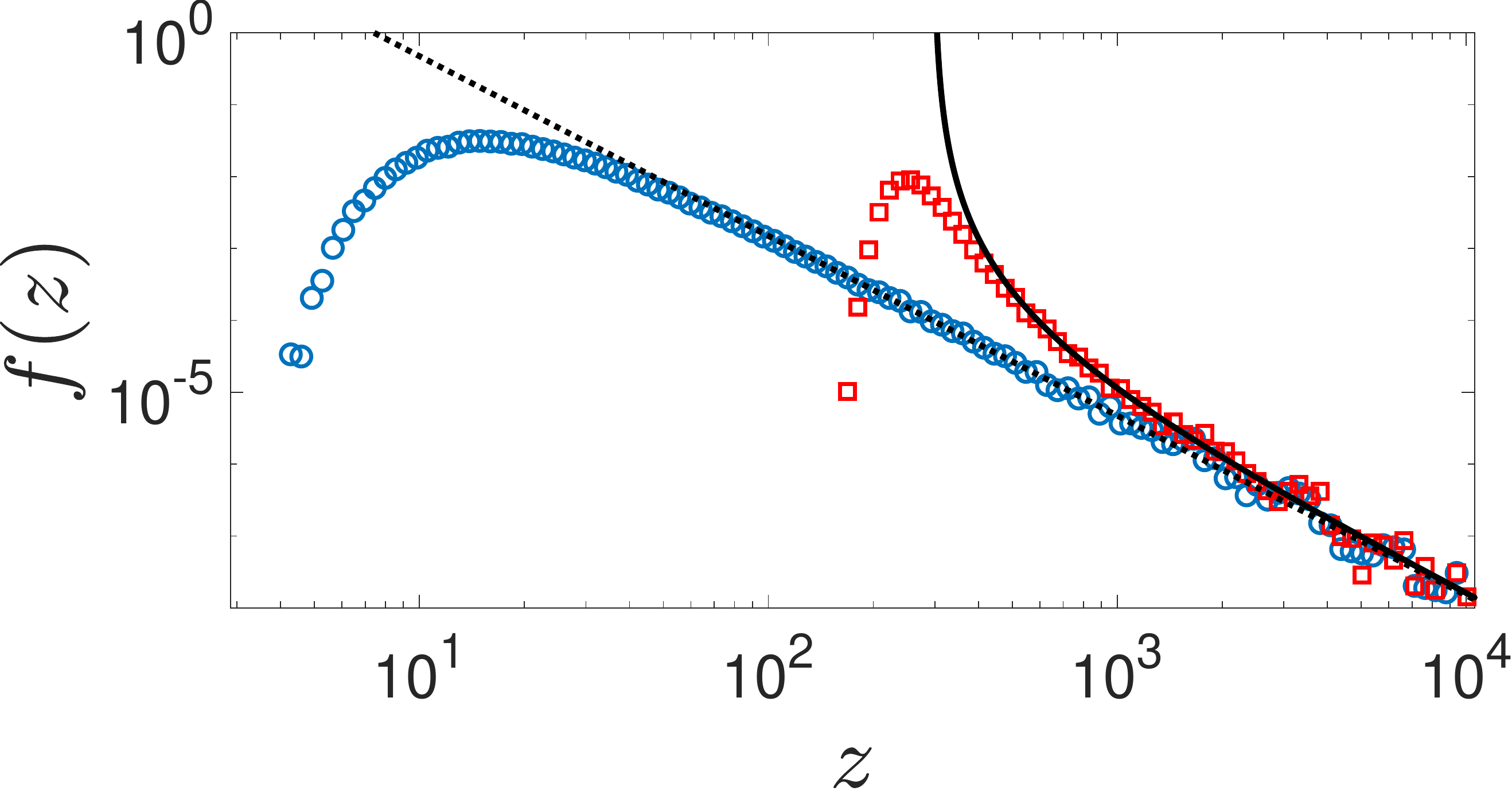}

\caption{Histograms of the IID maximum $\delta_\text{max}$ (blue) and the IID sum $x_N$ (red) compared with the IID BJP $N\alpha z^{-1-\alpha}$ (dotted line) and the correction $N\alpha(z-C)^{-1-\alpha}$ with $C=(N-1)\langle \delta_i \rangle$, see Eq.~(\ref{correction_ansatz}) and (\ref{final_theo_correct}). The IID random variables follow the Pareto PDF with $\alpha=1.5$. We used $N=10^2$ and $10^5$ realizations.\label{fig:plot_app_01}}
\end{center}\end{figure}

The random walk is defined by the sum $x_N=\sum_{i=1}^N \delta_i$. Now we assume that the maximum happened at $b$, i.e. $\delta_\text{max}=\delta_b$, is very large. Then we replace all remaining random variables $\delta_i$, $i \neq b$, by the mean $\langle \delta_i \rangle$. This replacement neglects the fluctuations of the remaining variables due to the dominating large value of the maximum. We get
\begin{equation}\label{correction_ansatz}
x_N \stackrel{d}{\sim} \delta_\text{max} + (N-1) \langle \delta_i \rangle
\end{equation}
or similarly $x_N \stackrel{d}{\sim} \delta_\text{max} +  \langle x_{N-1} \rangle$, see Fig.~\ref{fig:plot_app_01}. \\ \\

We can derive this relationship also from the expansion of the PDFs. The sum PDF is the convolution $f_{x_N}(z)=(f\ast\ldots\ast f)^{(N)}(z)$ which is in Laplace space the product $\hat{f}_{x_N}(s)=[\hat{f}_{\delta_i}(s)]^N$. The Laplace transform of the random variable PDF is $\hat{f}_{\delta_i}(s)=\alpha \Gamma(-\alpha)s^\alpha+1-\langle \delta_i \rangle s + \mathcal{O}(s^2)$. In order to get the tail of $f_{x_N}(z)$ we only need the lowest order non-integer exponents of $s$ in the Laplace transform. Hence, we use the small $s$ expansion $\hat{f}_{x_N}(s) \sim N \alpha \Gamma(-\alpha) s^\alpha -N(N-1)\alpha\langle \delta_i \rangle \Gamma(-\alpha)s^{1+\alpha}$. Inverse Laplace transform gives 
\begin{equation}\label{correction_ansatz2}
f_{x_N}(z)\sim N\alpha z^{-1-\alpha}+N(N-1)\alpha(1+\alpha)\langle \delta_i \rangle z^{-2-\alpha}.
\end{equation}
Now we need the maximum PDF which is $f_{\delta_\text{max}}(z)=Nf_{\delta_i}(z)[F_{\delta_i}(z)]^{N-1}\sim Nf_{\delta_i}(z) = N\alpha z^{-1-\alpha}$. In order to get the correction of Eq.~(\ref{correction_ansatz}), we make the ansatz for the variable transform $\delta_\text{max}+C$. The PDF is $f_{\delta_\text{max}+C}(z)=f_{\delta_\text{max}}(z-C)$. The large argument behaviour $N\alpha (z-C)^{-1-\alpha}$ with large $z-C$ is
\begin{equation}\label{correction_ansatz3}
f_{\delta_\text{max}+C}(z) \sim N\alpha z^{-1-\alpha}+N\alpha(1+\alpha)C z^{-2-\alpha}.
\end{equation}
Comparison between Eq.~(\ref{correction_ansatz2}) and (\ref{correction_ansatz3}) yields
\begin{equation}
C=(N-1)\langle \delta_i \rangle
\end{equation}
and therefore shows Eq.~(\ref{correction_ansatz}). In terms of the PDFs, the correction yields
\begin{equation}\label{final_theo_correct}
f_{x_N}(z) \sim f_{\delta_\text{max}}(z-C).
\end{equation} 
Summarized, we learn from this IID case that Eq.~(\ref{correction_ansatz}) can be found quite easily by the theme: Neglect the fluctuations of the remaining variables (which are not the maximum) and replace their values by the mean.

\subsection{Correlated random walk}

\begin{figure}[t]\begin{center}
\includegraphics[width=0.22\textwidth]{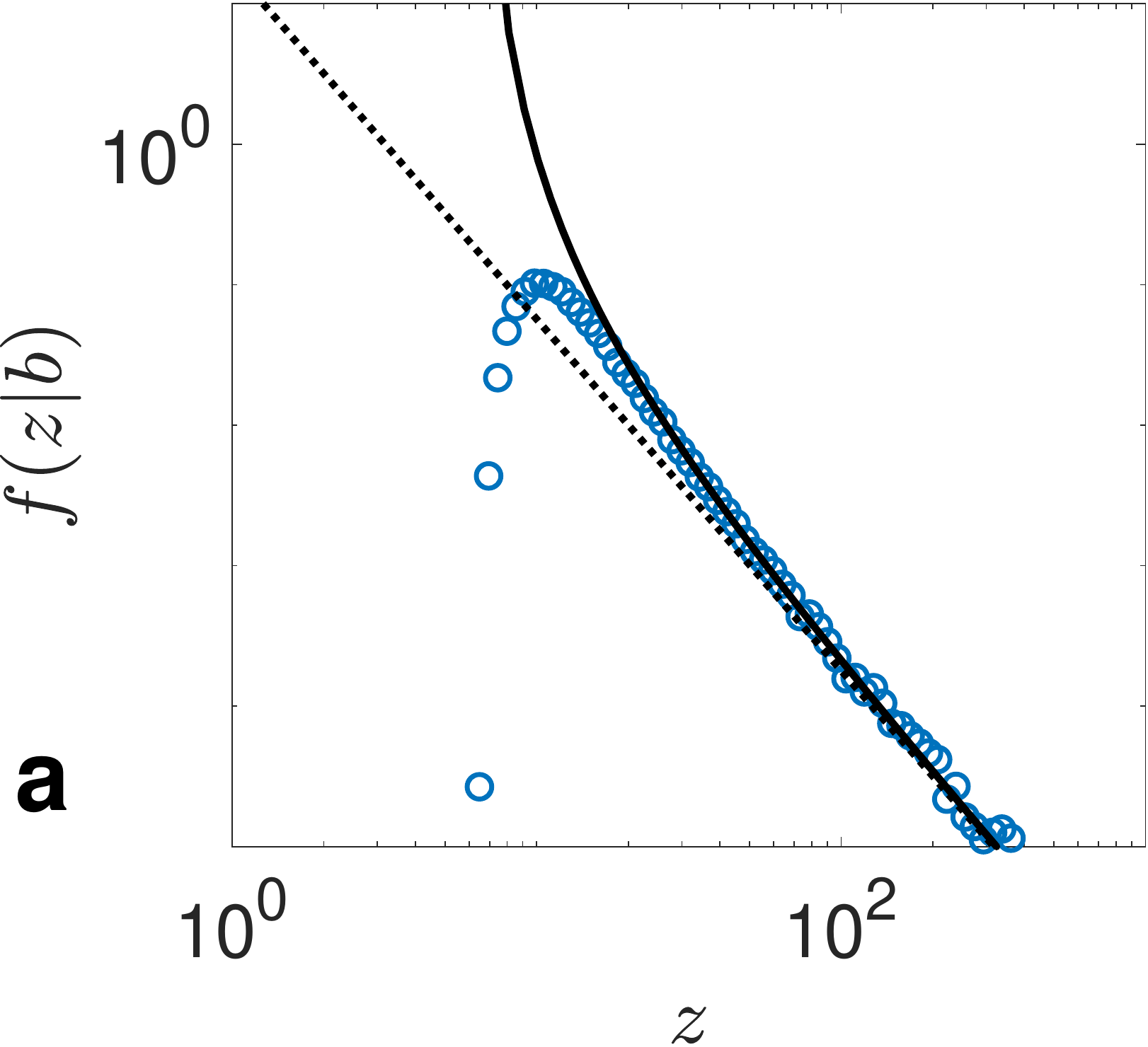}
\includegraphics[width=0.22\textwidth]{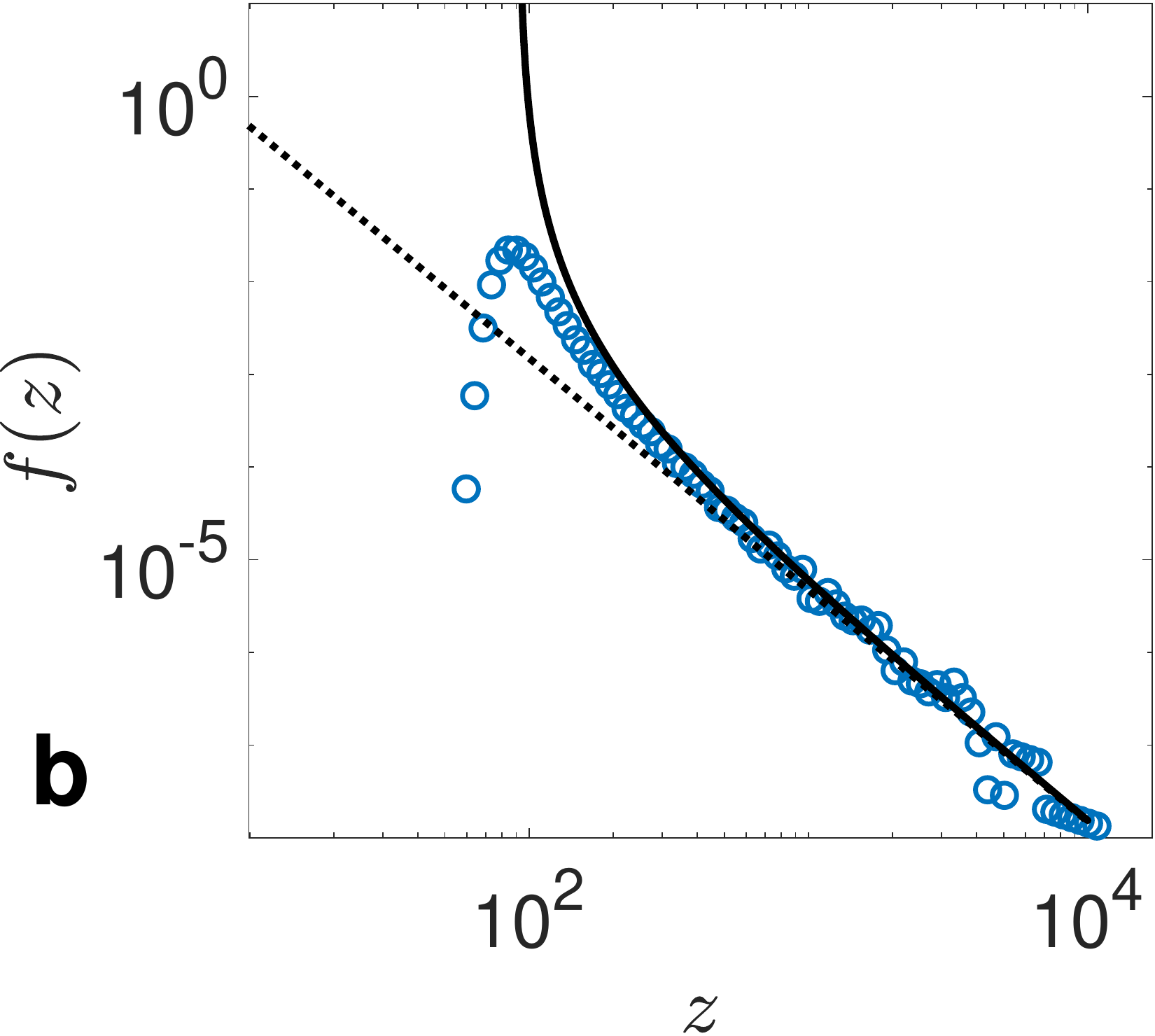}
\caption{\textbf{a} Histogram of $\tilde{\delta}_\text{max}$ (blue) conditioned on $b=4$ compared with the large $z$ behaviour $\alpha z^{-1-\alpha}/\Phi(b)$ (dotted line) and the correction of Eq.~(\ref{dd1}) (solid line). The $\delta_i$ follow the Pareto PDF with $\alpha=1.5$. The memory kernel is algebraical with $\beta=0.3$. We used $N=5$ and $10^6$ realizations. \textbf{b} Histogram of $\tilde{\delta}_\text{max}$ (blue) compared with the large $z$ behaviour $N\alpha z^{-1-\alpha}$ (dotted line) and the correction of Eq.~(\ref{dd2}) (solid line). The IID random variables follow the Pareto PDF with $\alpha=1.5$. The memory kernel is algebraical with $\beta=0.3$. We used $N=10^2$ and $10^5$ realizations.\label{fig:plot_app_02-03}}
\end{center}\end{figure}

\begin{figure}[t]\begin{center}
\includegraphics[width=0.22\textwidth]{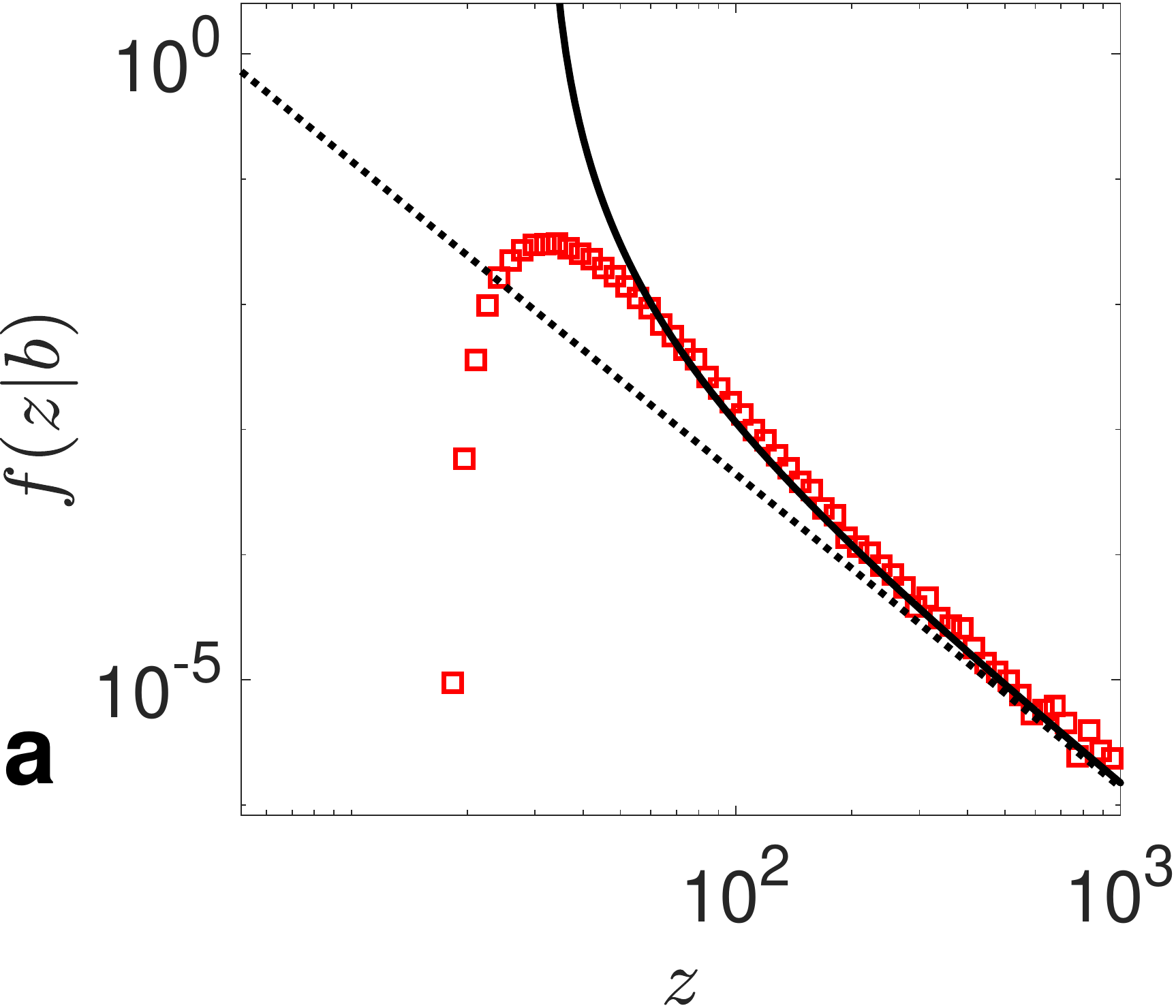}
\includegraphics[width=0.22\textwidth]{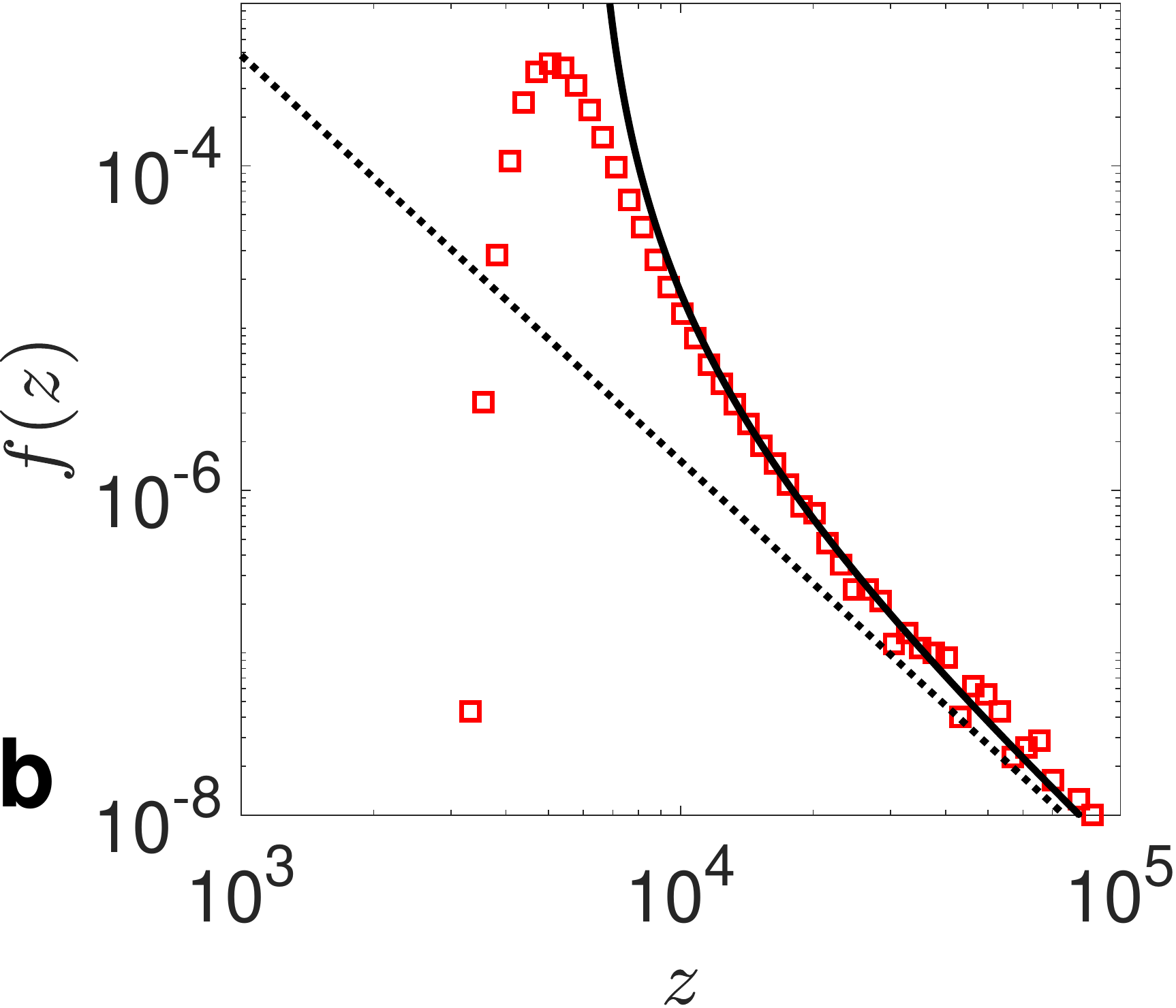}
\caption{\textbf{a} Histogram of $\tilde{x}_N$ (red) conditioned on $b=4$ compared with the large $z$ behaviour $(W_{N-b})^\alpha\alpha z^{-1-\alpha}/\Phi(b)$ (dotted line) and the correction of Eq.~(\ref{dd3}) (solid line). The $\delta_i$ follow the Pareto PDF with $\alpha=1.5$. The memory kernel is algebraical with $\beta=0.3$. We used $N=5$ and $10^6$ realizations. \textbf{b} Histogram of $\tilde{x}_N$ (red) compared with the large $z$ behaviour $\tilde{\gamma}_N\alpha z^{-1-\alpha}$ (dotted line) and the correction of Eq.~(\ref{dd4}) (solid line). The IID random variables follow the Pareto PDF with $\alpha=1.5$. The memory kernel is algebraical with $\beta=0.3$. We used $N=10^2$ and $10^5$ realizations.\label{fig:plot_app_04-05}}
\end{center}\end{figure}

We consider the correlated random walk model and start with the maximum $\tilde{\delta}_\text{max}$. We found the relationship $\tilde{\delta}_\text{max} \stackrel{d}{\sim} \delta_\text{max}$, see Eq.~(\ref{bjpmaximum}). Also here we can find a correction similar to the IID case of Eq.~(\ref{correction_ansatz}). Per definition the correlated increment is the weighted sum $\tilde{\delta}_i=\sum_{j=1}^i M_{i-j} \delta_j$, see Eq.~(\ref{inccorr}). We condition the appearance of the maximum $\tilde{\delta}_\text{max}$ at the step number $b$. Still per definition, it is $\tilde{\delta}_b=\delta_b+\sum_{j=1}^{b-1} M_{b-j} \delta_j$. When the maximum is very large, we replace the remaining variables by their mean (remember we assume $\alpha>1$ and Pareto). We get 
\begin{equation}
\tilde{\delta}_b \stackrel{d}{\sim}\delta_b+ \langle \delta_i \rangle \sum_{j=1}^{b-1} M_{b-j} .
\end{equation}
The IID maximum $\delta_b$ is conditioned on the occurrence of $\tilde{\delta}_\text{max}$. So $\delta_b$ follows the conditional PDF $f_{\delta_\text{max}|b}(z|b)\sim \alpha z^{-1-\alpha}/\Phi(b)$ where $\Phi(b)$ is the probability that $\tilde{\delta}_\text{max}$ happens at $b$, see Eq.~(\ref{condi}). We get the corrected conditional maximum PDF
\begin{equation}\label{dd1}
f_{\tilde{\delta}_\text{max}|b}(z|b)\sim \frac{1}{\Phi(b)} \alpha \left[z- \langle \delta_i \rangle \sum_{j=1}^{b-1} M_{b-j}\right]^{-1-\alpha},
\end{equation}
see Fig.~\ref{fig:plot_app_02-03}. The unconditional maximum PDF is obtained by summing over all $b$ with the weight $\Phi(b)$, see Eq.~(\ref{condi}). We get
\begin{equation}\label{dd2}
f_{\tilde{\delta}_\text{max}}(z)\sim N \alpha \left[z- \langle \delta_i \rangle \sum_{b=1}^N \left(\Phi(b)\sum_{j=1}^{b-1} M_{b-j}\right)\right]^{-1-\alpha},
\end{equation}
see Fig.~\ref{fig:plot_app_02-03}. \\ \\

Now we describe the corrections for the correlated sum PDF. We repeat the just presented ansatz. Per definition, it is $\tilde{x}_N = \sum_{k=1}^N W_{N-k}\delta_k$, see Eq.~(\ref{procorr}). We take out the $b$-th IID increment $\tilde{x}_N = W_{N-b} \delta_b +  \sum_{k=1,k \neq b}^N W_{N-k} \delta_k$. We assume that the big jump $\tilde{\delta}_\text{max}|b$ happens at $b$ and that it is also large so that $\tilde{\delta}_\text{max}|b \stackrel{d}{\sim} \delta_b$. Note that the IID big jump $\delta_b$ is conditioned on the step number of $\tilde{\delta}_\text{max}|b$. We replace the remaining increments by their mean and get
\begin{equation}\label{dd3}
\tilde{x}_N|b \stackrel{d}{\sim} W_{N-b} \delta_b + \langle \delta_i \rangle \sum_{k=1,k \neq b}^N W_{N-k}.
\end{equation}
Since $\delta_b \stackrel{d}{\sim}\alpha z^{-1-\alpha}/\Phi(b)$, we find the correction for the conditional BJP
\begin{equation}\label{dd4}
f_{\tilde{x}_N|b}(z|b) \sim \frac{(W_{N-b})^\alpha}{\Phi(b)} \alpha \left[z- \langle \delta_i \rangle \sum_{k=1,k \neq b}^N W_{N-k}\right]^{-1-\alpha},
\end{equation}
see Fig.~\ref{fig:plot_app_04-05}. The corrections for the unconditional BJP are obtained by summing over all $b$ with the weight $\Phi(b)$, see Eq.~(\ref{condi}). We get
\begin{equation}
f_{\tilde{x}_N}(z) \sim \tilde{\gamma}_N \alpha \left[z- \langle \delta_i \rangle \sum_{b=1}^N\left(\Phi(b) \sum_{k=1,k \neq b}^N W_{N-k}\right)\right]^{-1-\alpha},
\end{equation}
see Fig.~\ref{fig:plot_app_04-05}.

\section{Distribution of $\bm{\tilde{x}_N}$ in the large $N$ limit}\label{sec:applargenpdf}
\begin{figure}\begin{center}
\includegraphics[width=0.43\textwidth]{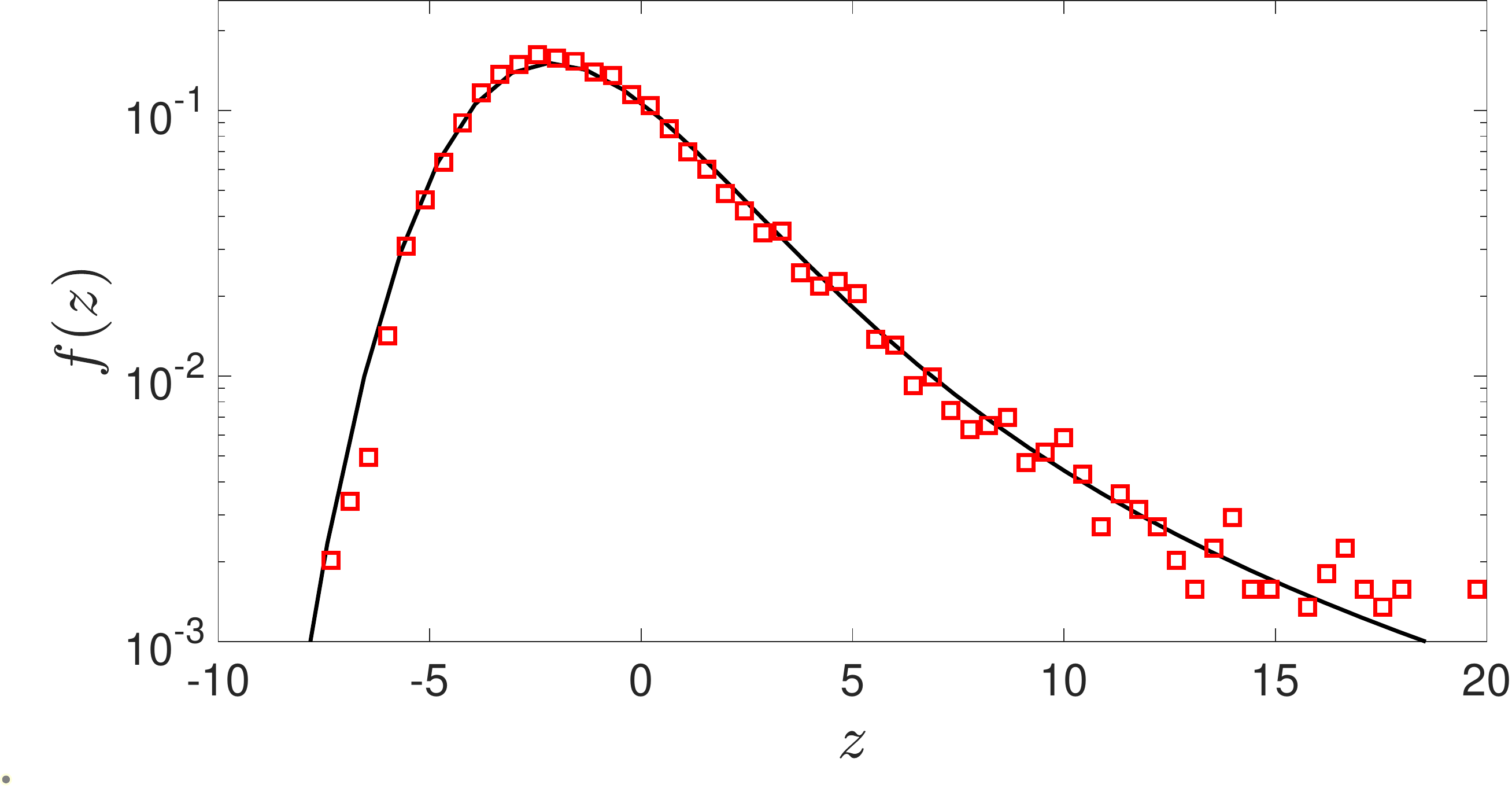}
\caption{Histogram for $(\tilde{\gamma}_N)^{-1/\alpha}(\tilde{x}_N-\mu)$ (red squares) compared with the limiting PDF $L_{\alpha,\kappa,c/\tilde{\gamma}_N,0}(z)$ (black line) where $N=10^4$, see the explanation for Eq.~(\ref{rescaledcharact}). The $\delta_i$ follow the Pareto PDF with $\alpha=3/2$. The memory kernel is exponential with the parameter $m=0.6$. The number of realizations is $10^4$.  \label{fig:plot_17}}
\end{center}\end{figure}

We calculate the large $N$ limit of the correlated sum PDF $f_{\tilde{x}_N}(z)$ where the uncorrelated increments follow the Pareto PDF $f_{\delta_i}(z) = \alpha z^{-1-\alpha}\Theta(z-1)$ with the Heaviside step function $\Theta(z-1)=1$ for $z \ge 1$ and $\Theta(z-1)=1$ for $z < 0$.\\ \\
We use the characteristic function which is the product
\begin{equation}
\varphi_{\tilde{x}_N}(k) = \prod_{i=1}^N\varphi_{\delta_i}(W_{N-i}k)
\end{equation}
First, we calculate the characteristic function of the uncorrelated increments
\begin{equation}\begin{split}
\varphi_{\delta_i}(k) &= \int\limits_{-\infty}^\infty e^{ikz}f_{\delta_i}(z)\mathrm{d}z \\
&=\int\limits_{-\infty}^\infty \text{cos}(kz)f_{\delta_i}(z)\mathrm{d}z + i \int\limits_{-\infty}^\infty \text{sin}(kz)f_{\delta_i}(z)\mathrm{d}z \\
&= \alpha\Gamma(-\alpha) |k|^\alpha \text{cos}\left(\frac{\pi\alpha}{2}\right)\\
&+ {}_pF_q \left(\left\lbrace - \frac{\alpha}{2} \right\rbrace , \left\lbrace\frac{1}{2},1-\frac{\alpha}{2} \right\rbrace,-\frac{k^2}{2} \right)\\
&+i\Bigg[-\alpha\Gamma(-\alpha) |k|^{-1+\alpha} \text{sin}\left(\frac{\pi\alpha}{2}\right)\\
&+\frac{\alpha}{\alpha-1}k {}_pF_q \left(\left\lbrace \frac{1}{2}- \frac{\alpha}{2} \right\rbrace , \left\lbrace\frac{3}{2},\frac{3}{2}-\frac{\alpha}{2} \right\rbrace,-\frac{k^2}{4} \right) \Bigg].
\end{split}\end{equation}
The last step can be checked with Mathematica using $f_{\delta_i}(z) = \alpha z^{-1-\alpha}\Theta(z-1)$. The small $k$ expansion is
\begin{equation}\begin{split}
\varphi_{\delta_i}(k) &\sim \alpha \Gamma(-\alpha)\text{cos}\left(\frac{\pi\alpha}{2}\right)|k|^\alpha \left[1-i\text{sign}(k) \text{tan}\left( \frac{\pi\alpha}{2}\right) \right]\\
&+ 1+i\frac{\alpha}{\alpha-1}k+ \mathcal{O}(k^2)
\end{split}\end{equation}
Therefore for large $N$ we get
\begin{equation}
\varphi_{\tilde{x}_N}(k) \sim \text{exp}\left[ik\mu - c |k|^\alpha \left(1-i\beta\text{sign}(k)\text{tan}\left(\frac{\pi\alpha}{2}\right)\right)\right]
\end{equation}
with
\begin{equation}\begin{split}
\beta &=1, \\
c &=-\alpha\Gamma(-\alpha) \text{cos}\left(\frac{\pi\alpha}{2}\right) \sum_{i=1}^N (W_{N-i})^\alpha,\\
\mu &=
\begin{cases}
0 & \text{ for } \alpha\in(0,1),\\
\frac{\alpha}{\alpha-1}\sum_{i=1}^N W_{N-i} & \text{ for } \alpha\in(1,2).
\end{cases}
\end{split}\end{equation}
Finally, the characteristic function of the shifted and rescaled correlated sum is
\begin{equation}\label{rescaledcharact}
\varphi_{A_N(\tilde{x}_N-B_N)}(k) = e^{-iA_NB_Nk}\prod_{i=1}^N\varphi_{\delta_i}(A_NW_{N-i}k).
\end{equation}
$A_N$ and $B_N$ can be chosen such that the limiting distribution is independent of $N$. One finds $A_N=(\tilde{\gamma}_N)^{-1/\alpha}$ and $B_N=\mu$ so that the rescaled random variable converges for large $N$ to $L_{\alpha,\kappa,c/\tilde{\gamma}_N,0}(z)$, see Fig.~\ref{fig:plot_17}.

\section{Scaling factor $\bm{\tilde{\gamma}_N}$ in the large $\bm{N}$ limit}\label{sec:scaling}
Here, we calculate the large $N$ limit of the scaling factor defined in Eq.~(\ref{scalingfa}) as 
\begin{equation}
\tilde{\gamma}_N = \sum_{k=1}^N (W_{N-k})^\alpha.
\end{equation}
First, we begin with the exponential memory kernel. From Eq.~(\ref{anotherlarge}) we know that the large $N$ behaviour of the weight is $W_{N-k} \sim 1/(1-m)$. Hence, we can  conclude
\begin{equation}
\tilde{\gamma}_N \sim \frac{1}{(1-m)^\alpha} N.
\end{equation}
Secondly, for the algebraical memory kernel we find the large $N$ limit of the scaling factor
\begin{equation}
\tilde{\gamma}_N = \sum_{k=1}^N \left( \sum_{j=0}^{N-k}(j+1)^{-\beta}\right)^\alpha
\end{equation}
by the following arguments. The inner sum behaves asymptotically as
\begin{equation}
\sum_{j=0}^{N-k}(j+1)^{-\beta} \sim \int_0^{n-k}(j+1)^{-\beta} \mathrm{d}j \sim \frac{(N-k+1)^{1-\beta}}{1-\beta}.
\end{equation}
The outer sum behaves asymptotically as the integral
\begin{equation}
\sum_{k=1}^N \left( \frac{(N-k+1)^{1-\beta}}{1-\beta}\right)^\alpha \sim \int_1^N \left( \frac{(N-k+1)^{1-\beta}}{1-\beta}\right)^\alpha \mathrm{d}k
\end{equation}
which can be calculated to 
\begin{equation}
\tilde{\gamma}_N \sim \frac{1}{(1-\beta)^\alpha} \frac{N^{1+\alpha(1-\beta)}}{1+\alpha(1-\beta)}
\end{equation}
with the exponent range $1<1+\alpha(1-\beta)<3$.

\section{Maximum CDF}\label{sec:appmax}

The maximum PDF of the correlated increments follows the same tail as the maximum PDF of the uncorrelated increments
\begin{equation}\label{tailmaxi}
f_{\tilde{\delta}_\text{max}}(z) \sim N A z^{-1-\alpha},
\end{equation}
i.e it is independent of the correlations. We derive this result by rewriting the definition of the maximum PDF and finding a formula suitable for a proof by induction. \\ \\

The maximum PDF is $f_{\tilde{\delta}_\text{max}}(z) = \mathrm{d}/\mathrm{d}z F_{\tilde{\delta}_\text{max}}(z)$ with the cumulative distribution function (CDF) $F_{\tilde{\delta}_\text{max}}(z)=\text{Prob}(\tilde{\delta}_1 \le z ,\ldots , \tilde{\delta}_N \le z)$ which is 
\begin{equation}\begin{split}
F_{\tilde{\delta}_\text{max}}(z) =  \int\limits_{-\infty}^z \mathrm{d}z_1\ldots \int\limits_{-\infty}^z  \mathrm{d}z_N f_{\tilde{\delta}_1,\ldots,\tilde{\delta}_N}(z_1,\ldots,z_N).
\end{split}\end{equation}
$f_{\tilde{\delta}_1,\ldots,\tilde{\delta}_N}(z_1,\ldots,z_N)$ is the joint PDF of the correlated increments $(\tilde{\delta}_1,\ldots,\tilde{\delta}_N)$. We rewrite this joint PDF using the chain rule of probability so that
\begin{equation}\label{chain}
f_{\tilde{\delta}_1,\ldots,\tilde{\delta}_N}(z_1,\ldots,z_N) = \prod_{i=1}^N f_{\tilde{\delta}_i|\tilde{\delta}_1,\ldots,\tilde{\delta}_{i-1}}(z_i|z_1,\ldots,z_{i-1})
\end{equation}
Note that for $i=1$ the right hand side gives $f_{\tilde{\delta}_1}(z_1)$. We can simplify these conditional PDFs. When the first $i-1$ uncorrelated increments $\delta_1,\ldots,\delta_{i-1}$ are given, then $i$th correlated increment is the sum of a constant and the $i$-th uncorrelated increment $\tilde{\delta}_i = \sum_{j=1}^{i-1}M_{i-j}\delta_j + \delta_i$, see Eq.~(\ref{inccorr}). Therefore, the conditional PDFs of Eq.~(\ref{chain}) are the single PDFs with shifted argument
\begin{equation}
f_{\tilde{\delta}_i|\tilde{\delta}_1,\ldots,\tilde{\delta}_{i-1}}(z_i|z_1,\ldots,z_{i-1}) = f_{\delta_i}\left(z_i-\sum_{j=1}^{i-1}M_{i-j}z_j \right).
\end{equation}
Hence, we get the correlated maximum CDF in a suitable form
\begin{equation}\label{suitable}
F_{\tilde{\delta}_\text{max}}(z) = \ \int\limits_{-\infty}^z\mathrm{d} z_1 \ldots \int\limits_{-\infty}^z \mathrm{d}z_N \prod_{i=1}^N f_{\delta_i}\left(z_i-\sum_{j=1}^{i-1}M_{i-j}z_j \right).
\end{equation}
Now we approximate this formula and find an expression predestinated for a proof by induction to show Eq.~(\ref{tailmaxi}). \\ \\

We approximate the first inner integral over $z_N$, namely
\begin{equation}\begin{split}\label{cdfapr}
&\int\limits_{-\infty}^z \mathrm{d}z_N f_{\delta_N} \left( z_N - \sum_{j=1}^{N-1}M_{N-1}z_j\right) \\
&= F_{\delta_N} \left( z - \sum_{j=1}^{N-1}M_{N-1}z_j\right)\sim F_{\delta_N}(z).
\end{split}\end{equation}
The approximation in the last step is due the binomial theorem for large values of $z$. The remaining $N-1$ integrals over $z_1,\ldots,z_{N-1}$ in Eq.~(\ref{suitable}) are exactly the maximum CDF for $N-1$ variables $F_{\tilde{\delta}_\text{max}}(z;N-1)$. We added the number of variables $N-1$ into the notation and will continue with this notation for the following formulas. Therefore the maximum CDF Eq.~(\ref{suitable}) behaves as
\begin{equation}\label{cdffinal}
F_{\tilde{\delta}_\text{max}}(z;N) \sim F_{\tilde{\delta}_\text{max}}(z;N-1) F_{\delta_N}(z).
\end{equation}

It is important to mention that we implied the condition $M_{N-j}<1$ for the validity of this formula. We explain it for $N=2$. The CDF $F_{\delta_2}(z-M_1z_1)$ in Eq.~(\ref{cdfapr}) requires $z-M_1z_1>1$ which yields $z_1<(z-1)/M_1 \sim z/M_1$ for large $z$. Only for $M_1<1$ it is $z/M_1>z$ so that the integral over $z_1$ in Eq.~(\ref{suitable}) goes until $z$ and not only until $z/M_1$. And this gives Eq.~(\ref{cdffinal}).\\ \\

The maximum PDF is the derivative of the maximum CDF Eq.~(\ref{cdffinal}) and behaves as
\begin{equation}
f_{\tilde{\delta}_\text{max}}(z;N) \sim f_{\tilde{\delta}_\text{max}}(z;N-1) F_{\delta_N}(z)+F_{\tilde{\delta}_\text{max}}(z;N-1) f_{\delta_N}(z).
\end{equation}
Furthermore, we can use here that both CDFs are about $1$ for large $z$ so that
\begin{equation}\label{induction}
f_{\tilde{\delta}_\text{max}}(z;N) \sim f_{\tilde{\delta}_\text{max}}(z;N-1)+ f_{\delta_N}(z).
\end{equation}
We finally found the formula suitable for a proof by induction to show the tail behaviour Eq.~(\ref{tailmaxi}). Starting with $N=2$ we can directly get $f_{\tilde{\delta}_\text{max}}(z;2) \sim 2 A z^{-1-\alpha}$. Furthermore, assuming the scaling $f_{\tilde{\delta}_\text{max}}(z;N) \sim NA z^{-1-\alpha}$ of Eq.~(\ref{tailmaxi}) is correct, we see from Eq.~(\ref{induction}) that $f_{\tilde{\delta}_\text{max}}(z;N+1) \sim (N+1)A z^{-1-\alpha}$. Thus, the scaling Eq.~(\ref{tailmaxi}) is indeed correct. 

\subsection{Pareto IID random variables for $N=2$}
\label{sec:pareto}
We calculate the maximum PDF $f_{\tilde{\delta}_\text{max}}(z)$ for $N=2$ when the IID random variables follow the Pareto PDF $f_{\delta_i}(z)=\alpha z^{-1-\alpha}$ with $z>1$. We use $m=M_1$. From Eq.~(\ref{suitable}), we get
\begin{equation}\begin{split}\label{ntwo}
&f_{\tilde{\delta}_\text{max}}(z) = \frac{\mathrm{d}}{\mathrm{d} z} F_{\tilde{\delta}_\text{max}}(z) \\
&= \int\limits_{-\infty}^z f_{\delta_i}(z) f_{\delta_i}(z_2-mz) \mathrm{d}z_2 + \int\limits_{-\infty}^z f_{\delta_i}(z_1) f_{\delta_i}(z-mz_1) \mathrm{d}z_1.
\end{split}\end{equation}
The first integral gives
\begin{equation}\label{firstresulti}
\int\limits_{-\infty}^z f_{\delta_i}(z) f_{\delta_i}(z_2-mz) \mathrm{d}z_2 = f_{\delta_i}(z)F_{\delta_i}(z-mz).
\end{equation}
For the second integral, we derive the indefinite integral
\begin{equation}\label{indefini}
\int f_{\delta_i}(z_1) f_{\delta_i}(z-mz_1) \mathrm{d}z_1 = I(z_1,z).
\end{equation}
Mathematica gives us
\begin{equation}\begin{split}
I(z_1,z)&= \frac{\alpha}{1-\alpha} z^{-2} {z_1}^{-\alpha}(z-mz_1)^{-\alpha}\left(1-\frac{mz_1}{z}\right)^\alpha\\
&\times \Big[\alpha mz_1 \times {}_2F_1(1-\alpha,1+\alpha,2-\alpha,mz_1/m)\\
&+(\alpha-1)z \times {}_2F_1(-\alpha,\alpha,1-\alpha,mz_1/z)\Big]
\end{split}\end{equation}
where ${}_2F_1$ is the hypergeometric function. The integrand of Eq.~(\ref{indefini}) yields 3 different regions for $z$. Thus, we get the second integral as
\begin{equation}\begin{split}\label{secresulti}
&\int\limits_{-\infty}^z f_{\delta_i}(z_1) f_{\delta_i}(z-mz_1) \mathrm{d}z_1\\
&=
\begin{cases}
0 & , z \in(-\infty, 1+m),\\
I\left(\frac{z-1}{m},z \right)-I(1,z) & , z \in\left(1+m,\frac{1}{1-m}\right),\\
I(z,z)-I(z,1) & , z \in\left(\frac{1}{1-m},\infty\right).\\
\end{cases}
\end{split}\end{equation}
So we find the maximum PDF excatly as the sum of the first integral Eq.~(\ref{firstresulti}) and the second integral Eq.~(\ref{secresulti}). \\ \\

The large $z$ behaviour is the sum of the large $z$ behaviours of Eq.~(\ref{firstresulti}) and (\ref{secresulti}). The first integrals behaves as $f_{\delta_i}(z)=\alpha z^{-1-\alpha}$. The second integral gives also (after some calculations) the same behaviour $\alpha z^{-1-\alpha}$. So that we finally have
\begin{equation}
f_{\tilde{\delta}_\text{max}}(z)\sim 2 \alpha z^{-1-\alpha}.
\end{equation}
This is the same large $z$ behaviour as for the IID maximum $f_{\delta_\text{max}}(z)\sim 2\alpha z^{-1-\alpha}$. Therefore it shows $\tilde{\delta}_\text{max} \stackrel{d}{\sim} \delta_\text{max}$ for $N=2$ where we get the large $z$ behaviour from the exact expression of $f_{\tilde{\delta}_\text{max}}(z)$.

\subsection{Uniform IID random variables for $N=2$}\label{sec:uniform}
We calculate the maximum PDF $f_{\tilde{\delta}_\text{max}}(z)$ for $N=2$ when the IID random variables follow the uniform PDF on the interval $[0,1]$, i.e. $f_{\delta_i}(z)=\Theta(1-z)\Theta(z)$ where $\Theta$ is the Heaviside step function. We use $m=M_1$. From Eq.~(\ref{suitable}), we get
\begin{equation}\begin{split}\label{ntwo}
&f_{\tilde{\delta}_\text{max}}(z) = \frac{\mathrm{d}}{\mathrm{d} z} F_{\tilde{\delta}_\text{max}}(z) \\
&= \int\limits_{-\infty}^z f_{\delta_i}(z) f_{\delta_i}(z_2-mz) \mathrm{d}z_2 + \int\limits_{-\infty}^z f_{\delta_i}(z_1) f_{\delta_i}(z-mz_1) \mathrm{d}z_1.
\end{split}\end{equation}
The first integral gives
\begin{equation}
\int\limits_{-\infty}^z f_{\delta_i}(z) f_{\delta_i}(z_2-mz) \mathrm{d}z_2 = (1-m)z \Theta(z) \Theta(1-z).
\end{equation}
The second integral gives
\begin{equation}\begin{split}
&\int\limits_{-\infty}^z f_{\delta_i}(z_1) f_{\delta_i}(z-mz_1) \mathrm{d}z_1\\
&= z \Theta(z) \Theta(1-z) + \frac{1+m-z}{m}\Theta(z-1) \Theta(1+m-z).
\end{split}\end{equation}
Finally, the maximum PDF is
\begin{equation}
f_{\tilde{\delta}_\text{max}}(z) =
\begin{cases}
(2-m)z & \text{ for } 0<z<1,\\
(1+m-z)/m & \text{ for } 1<z<1+m
\end{cases}
\end{equation}
which we observe in Fig.~\ref{fig:plot_01-04}.
\end{appendix}

\bibliographystyle{prestyle}
\bibliography{bjpcorr_citations}

\end{document}